\documentclass[prl,aps,superscriptaddress,twocolumn]{revtex4-2}

\usepackage{subfigure}
\usepackage{xcolor}
\usepackage{graphicx}
\usepackage{dcolumn}
\usepackage{bm}
\usepackage{amssymb}
\usepackage{comment}
\usepackage{physics}

\usepackage[colorlinks=true]{hyperref}
\hypersetup{linkcolor=blue,citecolor=blue,urlcolor=blue}

\newcommand{\be}{\begin{equation}}
\newcommand{\ee}{\end{equation}}

\newcommand{\bea}{\begin{eqnarray}}
\newcommand{\eea}{\end{eqnarray}}
\newcommand{\ba}{\begin{align}}
\newcommand{\ea}{\end{align}}

\newcommand{\beq}{\begin{equation}}
\newcommand{\eeq}{\end{equation}}

\begin{document}

\title{Extracting central charge from ground-state overlaps of \\
spatially deformed Hamiltonians}

\author{Chen Bai}
\thanks{These two authors contributed equally.}
\affiliation{Kavli Institute for Theoretical Sciences, University of Chinese Academy of Sciences, Beijing 100190, China}

\author{Xinyu Sun}
\thanks{These two authors contributed equally.}
\affiliation{Institute for Advanced Study, Tsinghua University, Beijing 100084, China}

\author{Liang-Hong Mo}
\affiliation{Department of Physics, Princeton University, Princeton, New Jersey, 08544, USA}

\author{Hong-Hao Tu}
\email{h.tu@lmu.de}
\affiliation{Faculty of Physics and Arnold Sommerfeld Center for Theoretical Physics, Ludwig-Maximilians-Universit\"at M\"unchen, 80333 Munich, Germany}

\date{\today}

\begin{abstract}
We show that the conformal anomaly of a $(1+1)$-dimensional conformal field theory can be extracted directly from a ground-state wave-function overlap associated with a spatial conformal deformation. Focusing on the $q$-M\"obius deformation, we derive an exact overlap formula between the deformed and undeformed ground states, whose exponent directly encodes the central charge. Motivated by this result, we construct a lattice estimator based solely on ground-state overlaps and apply it to representative critical quantum chains and the gapless edge modes of a two-dimensional Chern insulator. Numerical results demonstrate that the resulting overlaps provide a simple and robust probe of the central charge in microscopic models. We further demonstrate that the deformed ground states retain universal geometric structures in their entanglement spectra and entanglement entropies. These results provide a simple wave-function-based route to probing conformal data in critical systems and topological edge modes.
\end{abstract}

\maketitle

\textit{Introduction}.
Gapless one-dimensional (1D) quantum systems, from critical lattice models to boundary modes of two-dimensional (2D) topological phases, are described at low energies by $(1+1)$-dimensional conformal field theories (CFTs)~\cite{Belavin1984}. Among their universal data, the central charge $c$ is especially fundamental: it is the Virasoro conformal anomaly~\cite{DiFrancesco1997}, fixes the universal Casimir term in finite-size energies~\cite{Bloete1986,Affleck1986}, governs the logarithmic scaling of entanglement entropy~\cite{Holzhey1994,Vidal2003,Calabrese2004}, and quantifies effective low-energy degrees of freedom along renormalization group flows~\cite{Zamolodchikov1986}. Extracting $c$ from a microscopic model is therefore a key diagnostic for identifying the infrared CFT of a critical system or gapless edge mode.

Existing lattice diagnostics of the central charge include finite-size spectral scaling~\cite{Cardy1984a,Bloete1986,Affleck1986}, entanglement entropy scaling~\cite{Vidal2003,Calabrese2004}, return amplitudes after quantum quenches~\cite{Cardy2014,Lapierre2025a,Lapierre2025b,MoLH2026}, Koo-Saleur-type lattice Virasoro constructions~\cite{Koo1993,Milsted2017}, and wave-function-overlap probes~\cite{ZouYJ2022a,LiuYH2023}. 
For gapless edges of 2D topological phases, related probes include momentum polarization~\cite{TuHH2013,Zaletel2013,Kobayashi2024} and modular-commutator approaches~\cite{KimIH2022a,KimIH2022b,ZouYJ2022b,FanRH2022}.
These methods often involve entanglement measurements, dynamical protocols, geometry-dependent constructions, or excited-state information. 
This motivates a simpler ground-state-overlap probe that directly captures the conformal anomaly.

Spatially deformed CFTs provide a natural setting for such overlap probes. 
When the Hamiltonian density is modulated by an envelope function associated with a conformal transformation, the resulting nonuniform Hamiltonian can remain analytically tractable. 
The M\"obius and sine-square deformations~\cite{Hikihara2011,Katsura2011,Maruyama2011} are well-known examples generated by the global conformal sector and closely related to the uniform theory obtained by radial quantization~\cite{Katsura2012,Tada2015,Okunishi2016,Tamura2017}. 
By contrast, higher-mode deformations, such as the $q$-M\"obius deformation~\cite{Miyata2024,BaiC2024}, involve non-global Virasoro generators~\cite{FanRH2019,HanB2020}. 
Because these generators are sensitive to the central extension, their ground-state overlaps can encode the conformal anomaly.

In this Letter, we show that the $q$-M\"obius deformation generates a universal ground-state overlap depending only on the central charge and deformation parameters, independent of system size and nonuniversal velocity. This yields a direct wave-function-based estimator of the central charge, potentially from a single sufficiently large system. 
We realize the deformation on the lattice by applying the same envelope function to local Hamiltonian terms. For representative critical chains, including the Ising, three-state Potts, spin-$\frac{1}{2}$ Heisenberg, and $\mathrm{SU}(3)$ Uimin-Lai-Sutherland chains, the finite-size estimates rapidly approach the expected CFT values.
We also show that, for long-range Hamiltonians with ground states expressible as chiral CFT correlators, such as the Haldane-Shastry model, the deformation can be implemented directly at the wave-function level.
We further uncover universal geometric structures in the entanglement properties of the deformed ground states and extend the construction to a 2D topological state, where deforming both edges or a single edge reproduces the expected total or chiral central charge, respectively.

\textit{$q$-M\"obius deformation}.
We consider a $(1+1)$-dimensional \textit{unitary} CFT on a circle of circumference $L$ with periodic boundary conditions. 
The $q$-M\"obius deformation is introduced by replacing the uniform Hamiltonian density with a spatially modulated one,
\begin{align}
    H_{q}(\theta)
    =
    \int_0^L \frac{\dd x}{2\pi}\,
    f_{q,\theta}(x)\left[T(x)+\bar{T}(x)\right],
\label{eq:CFT-Hq}
\end{align}
where $T(x)$ and $\bar{T}(x)$ denote the holomorphic and anti-holomorphic components of the stress tensor~\cite{DiFrancesco1997}. 
The envelope function is given by~\cite{Okunishi2016,FanRH2019}
\begin{align}
    f_{q,\theta}(x)
    =
    1-\tanh(2\theta)\cos\left(\frac{2q\pi x}{L}\right),
\label{eq:envelope-func}
\end{align}
where $q\in\mathbb{N}_+$ controls the modulation wave number and $\theta\in\mathbb{R}$ controls the deformation strength. 
At $\theta=0$, one has $f_{q,0}(x)=1$, and the undeformed uniform Hamiltonian is recovered. 
Hereafter, we denote it by $H$.

The special form of the envelope function makes the deformation analytically tractable. 
Indeed, we can express the stress tensor in Virasoro modes, with $L_n = \frac{L}{(2\pi)^2}\int_0^L \dd x\, T(x)e^{i 2\pi n x/L}+\frac{c}{24}\delta_{n,0}$ and similarly for $\bar L_n$, where $c$ is the central charge of the CFT.
In terms of these generators, the $q$-M\"obius Hamiltonian~\eqref{eq:CFT-Hq} can be written as
\begin{align}
  H_q(\theta) &= \frac{2\pi}{L}
    \left[L_0 -\frac{\tanh(2\theta)}{2}
        \left(L_q+L_{-q}\right) \right.  \nonumber \\
        &\phantom{=} \; \left. +\bar L_0
        -\frac{\tanh(2\theta)}{2}
        \left(\bar L_q+\bar L_{-q}\right)
        -\frac{c}{12} \right],
\end{align}
where the cosine modulation only couples the uniform Hamiltonian to the modes $L_{\pm q}$ and $\bar L_{\pm q}$. 
Thus, the $q$-M\"obius-deformed Hamiltonian is not a generic inhomogeneous Hamiltonian, but is generated by a finite conformal transformation. 
More precisely, its relation with the uniform Hamiltonian is given by
\begin{align}
 H_q(\theta) = \frac{1}{\cosh(2\theta)}
    U_q(\theta) H U_q^{\dagger}(\theta) - \frac{\pi c q^2}{6L}
    \left[1-\frac{1}{\cosh(2\theta)}\right],
\label{eq:Hq-unitary}
\end{align}
where $U_q(\theta) = e^{-\frac{\theta}{q}(L_q-L_{-q})}e^{-\frac{\theta}{q}(\bar L_q-\bar L_{-q})}$ is a unitary transformation; see the Supplemental Material (SM)~\cite{SM}. 
This relation shows that the deformation preserves the Virasoro structure of the CFT: it rescales the spectrum by $1/\cosh(2\theta)$ and shifts the energy by a constant.

For a unitary CFT, the ground state of the uniform Hamiltonian is the conformal vacuum $\ket{0}$. 
Equation~\eqref{eq:Hq-unitary} therefore implies that the ground state of the deformed Hamiltonian is obtained by acting on the vacuum with the same conformal transformation $\ket{0_q(\theta)}=U_q(\theta)\ket{0}$.
Rewriting this state yields
\begin{align}
    \ket{0_q(\theta)}
    =
    [\cosh\theta]^{-\frac{c(q^2-1)}{6q}}
    \exp\left[
        \frac{\tanh\theta}{q}
        \left(L_{-q}+\bar L_{-q}\right)
    \right]
    \ket{0},
\label{eq:CFT-deformed-GS}
\end{align}
which follows from disentangling the conformal transformation into normal-ordered form, as derived in the SM~\cite{SM}.
For $q=1$ or $\theta=0$, this state coincides with the uniform ground state as $L_{-1}\ket{0}=\bar L_{-1}\ket{0}=0$~\cite{Katsura2012}. 
For $q>1$ and $\theta\neq 0$, however, the deformed ground state is a nontrivial descendant coherent state built from the vacuum. 
Its overlap with the uniform ground state is
\begin{align}
    \langle 0 | 0_q(\theta)\rangle
    =
    [\cosh\theta]^{-\frac{c(q^2-1)}{6q}} .
\label{eq:CFT-Central-Charge}
\end{align}
Remarkably, this universal ground-state overlap directly encodes the central charge while remaining independent of both the system size and the nonuniversal velocity.

More generally, applying the same construction to a primary state $\ket{\phi}$ with scaling dimension $\Delta$ gives $\langle \phi|\phi_q(\theta)\rangle = [\cosh\theta]^{-\frac{2\Delta}{q}-\frac{c(q^2-1)}{6q}}$. 
Wave-function overlaps under $q$-M\"obius deformation therefore provide a direct probe of conformal data in the underlying CFT. While we focus below on extracting the central charge, we demonstrate in the SM~\cite{SM} that this approach accurately extracts the scaling dimensions in several paradigmatic examples.

We now devise a microscopic realization. Consider a 1D critical lattice Hamiltonian with periodic boundary conditions,
\begin{equation}
\tilde H = \sum_X h_X ,
\label{eq:lattice-H0}
\end{equation}
where each finite-range interaction $h_X$ is assigned to the center $X$ of its support. Taking an onsite term at site $j$ to be assigned to $X = j-\frac{1}{2}$, a term spanning sites $j$ to $j+r$ is assigned to $X=j+\frac{r-1}{2}$. The lattice $q$-M\"obius deformation is then
\begin{equation}
\tilde{H}_q(\theta)
= \sum_X f_{q,\theta}(X)h_X .
\label{eq:lattice-Hq}
\end{equation}
For fixed $q$, finite $\theta\neq0$, and finite interaction range, the envelope varies smoothly across each interaction as $L\to\infty$, as in the Koo-Saleur construction~\cite{Koo1993,Milsted2017}. In SM \cite{SM}, we apply this prescription to the critical cluster-Ising model, weighting each three-site term at its center.

For systems with long-range interactions, it is not obvious how to implement the $q$-M\"obius deformation at the Hamiltonian level. A useful class of examples is provided by lattice wave functions constructed as chiral correlators of CFTs~\cite{CiracSierra2010}, whose parent Hamiltonians generally contain long-range interactions~\cite{NielsenCiracSierra2011}. For such states, the $q$-M\"obius deformation can be implemented directly at the wave-function level through a conformal transformation of the lattice coordinates, from which a corresponding deformed parent Hamiltonian can be derived. In the SM~\cite{SM}, we present an explicit construction for the spin-1/2 Haldane-Shastry model~\cite{Haldane1988a,Shastry1988}.

\begin{figure}
    \centering
    \includegraphics[width=1.0\linewidth]{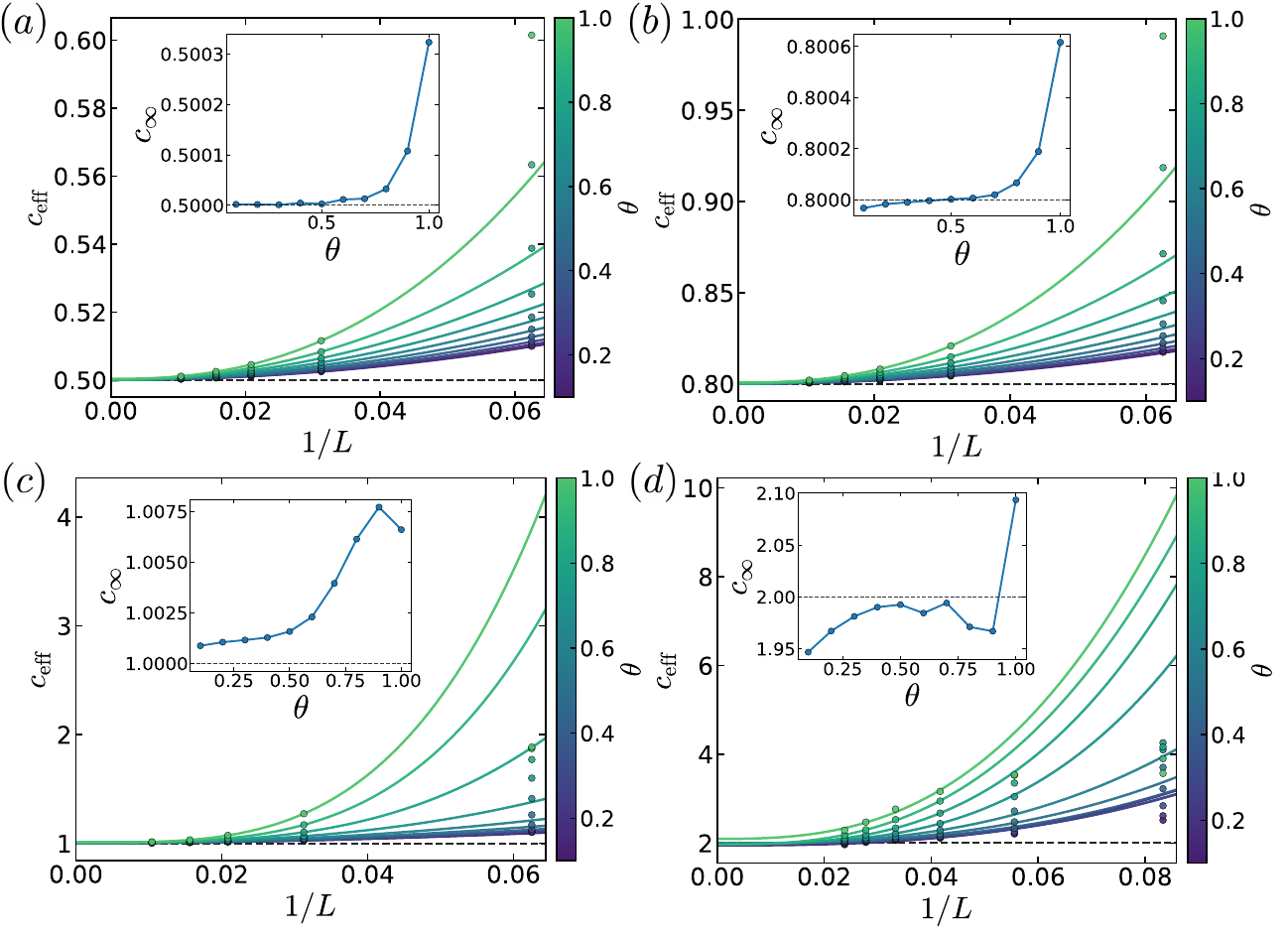}
    \caption{Finite-size extrapolation of the effective central charge for $q=2$ and different deformation parameters $\theta$ in four paradigmatic critical chains: (a) Ising, (b) three-state Potts, (c) spin-$\frac{1}{2}$ Heisenberg, and (d) SU(3) Uimin-Lai-Sutherland chains.
    Main panels show $c_{\mathrm{eff}}$ versus $1/L$, with colors indicating $\theta$. 
    Solid curves show the finite-size fits with $aL^{-p}+b$ based on the last four largest sizes, and insets show the extrapolated central charge $c_{\infty}$.
    Dashed lines indicate the expected central charges of the underlying CFTs. DMRG calculations are performed with bond dimension $\chi = 800$ for $L=16,32,48,64,96$ in (a)-(c), and $\chi = 1200$ for $L=12,18,24,30,36,42$ in (d).}
\label{fig:central charge of TFIM Potts XXZ SU3}
\end{figure}

Let $\ket{\Omega_0}$ and $\ket{\Omega_q(\theta)}$ be the ground states of the uniform and deformed lattice Hamiltonians [Eqs.~\eqref{eq:lattice-H0} and \eqref{eq:lattice-Hq}], respectively. 
The effective central charge is defined from their lattice ground-state overlap:
\begin{align}
    c_{\rm eff}(L;q,\theta)
    = -\frac{6q}{q^2-1}
    \frac{\ln \left|\langle\Omega_0|\Omega_q(\theta)\rangle\right|}{\ln[\cosh\theta]}.
\label{eq:Central-Charge-Lattice}
\end{align}
Using the density matrix renormalization group (DMRG) method~\cite{White1992,Schollwoeck2011}, we compute $c_{\rm eff}$ for representative critical chains, including the critical Ising, three-state Potts, spin-$\frac{1}{2}$ Heisenberg, and SU(3) Uimin-Lai-Sutherland chains~\cite{Uimin1970,LaiCK1974,Sutherland1975}. 
The uniform and deformed ground states are represented as matrix product states (MPSs), whose overlaps determine $c_{\rm eff}$. 
As shown in Fig.~\ref{fig:central charge of TFIM Potts XXZ SU3}, for $q=2$ and different $\theta$, $c_{\rm eff}$ approaches the expected CFT central charges with increasing $L$. For suitable deformation strengths, accurate estimates can already be obtained from a single sufficiently large system, while finite-size extrapolation can further refine these estimates.

\textit{Entanglement structure of deformed ground states}.
Beyond the universal overlap formula in Eq.~\eqref{eq:CFT-Central-Charge}, we ask how the $q$-M\"obius deformation modifies the internal structure of the ground state. 
Since the deformation is generated by a conformal transformation, its imprint should also appear in entanglement properties. 
As we now show, the deformed ground states exhibit remarkably simple and universal geometric structures in both the entanglement spectrum and entanglement entropy.

In a $(1+1)$-dimensional CFT, the ground state $\ket{0}$ is prepared by the Euclidean path integral on an infinite cylinder with spatial circumference $L$. The $q$-M\"obius ground state $\ket{0_q(\theta)}$ is obtained from $\ket{0}$ by the conformal transformation generated by $U_q(\theta)$. Equivalently, this transformation maps the original cylinder to one with effective circumference $L_{\rm eff}=L\cosh(2\theta)$. 
On the equal-time slice of the original cylinder, the map reduces to a spatial reparametrization $x\mapsto x^{\rm new}(x)$, given by
\begin{align}
    x^{\rm new}(x)=\frac{iL_{\rm eff}}{2\pi}\ln\left(\frac{\cosh\theta-\sinh\theta e^{i\frac{2q\pi x}{L}}}{\cosh\theta e^{i\frac{2q\pi x}{L}}-\sinh\theta }\right)^{\frac{1}{q}}.
\end{align}
Note that, as written, this function is multivalued in $x$. 
We choose the branch such that $x^{\rm new}(x)$ increases monotonically on $[0,L]$, with $x^{\rm new}(0)=0$ and $x^{\rm new}(L)=L_{\mathrm{eff}}$. 
Details are provided in the SM~\cite{SM}.

For a subsystem $A=[x_1,x_2]$, the reduced density operator $\rho_A$ of $\ket{0_q(\theta)}$ follows from applying the same conformal map to the vacuum reduced-density-matrix path integral.
This maps the replicated geometry to an annulus whose width is determined by the endpoint separation in the reparametrized coordinate and the local Jacobian factors at the entangling points. 
As shown in the SM~\cite{SM}, the resulting annulus width is
\begin{align}
W_A^{(q,\theta)}(x_1,x_2) &= \ln \left[\sin^2\left(
        \frac{\pi\,[x^{\text{new}}(x_1)-x^{\text{new}}(x_2)]}
             {L_{\rm eff}}
        \right)\right] \nonumber \\
    &\phantom{=} \; +\ln \left[
        f_{q,\theta}(x_1)f_{q,\theta}(x_2)
        \frac{L_{\rm eff}^2}{\pi^2\epsilon^2}
    \right],
\label{eq:qMobius-width}
\end{align}
where $\epsilon$ is the short-distance cutoff.

\begin{figure}
    \centering
    \includegraphics[width=1.0\linewidth]{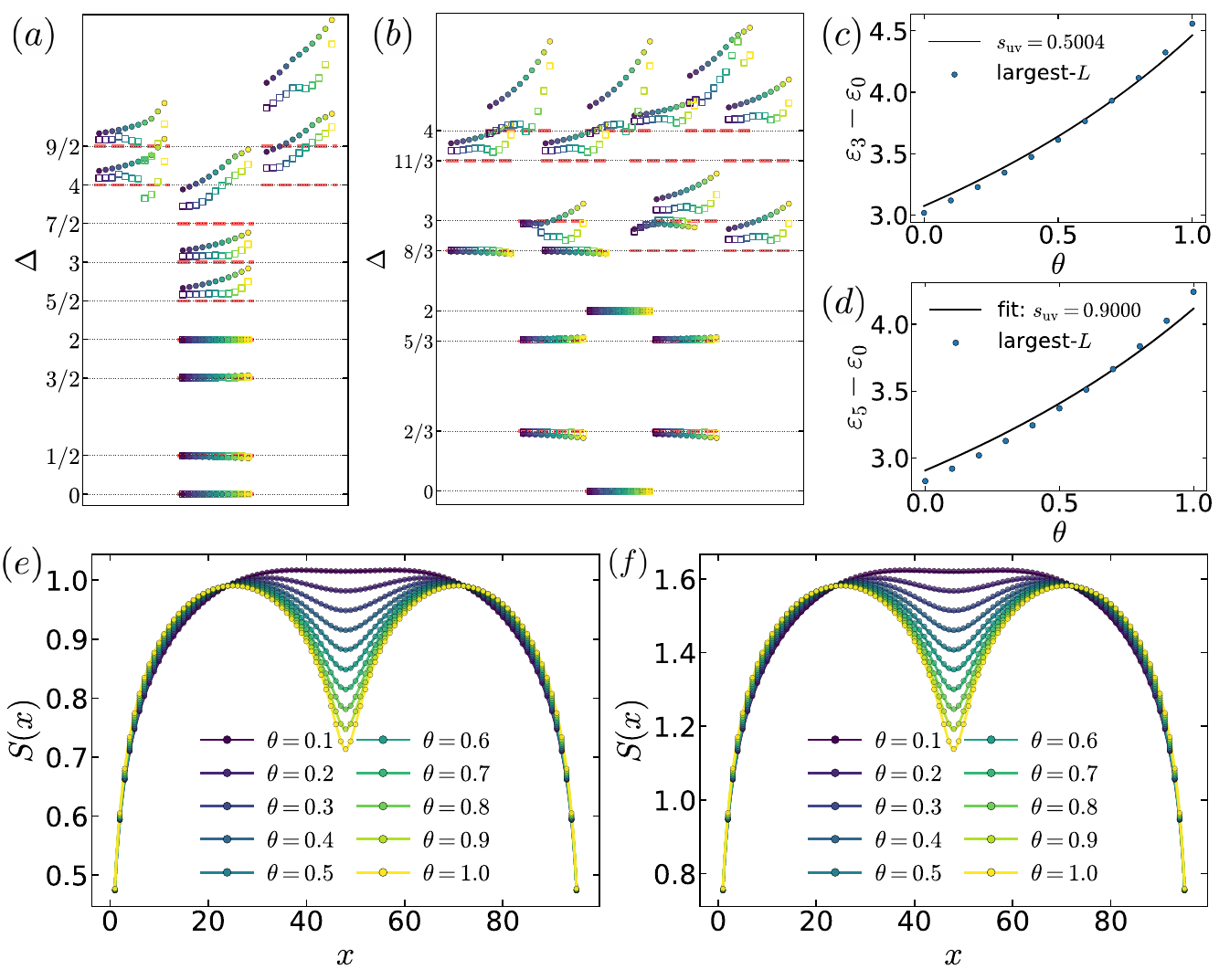}
    \caption{Entanglement spectrum and entanglement entropy ($S(x)=S_A(0,x)$) of the $q=2$ deformed Ising and three-state Potts chains. Panels (a)--(d) correspond to the half-chain subsystem $A = [0,L/2]$. (a,b) Rescaled entanglement spectra for the (a) Ising and (b) three-state Potts chains, normalized by setting the stress-tensor level to $\Delta=2$. Different colors correspond to different $\theta$ values, as indicated in panels (e) and (f). Circles represent the numerical data for $L=96$, squares show the thermodynamic-limit values extrapolated using the fitting form $aL^{-p}+b$ based on the last four largest sizes, and red dashed lines mark the BCFT predictions (free boundary conditions on both entangling points).
    (c,d) Spectral gaps used for rescaling as functions of $\theta$ for the (c) Ising and (d) three-state Potts models. 
    Blue dots show the largest-$L$ numerical data, while black curves show fits to Eq.~\eqref{eq:qMobius-ES-gap} with the nonuniversal ultraviolet (UV) constant $s_{\mathrm{uv}}$~\cite{UVcutoff}.
    (e,f) Entanglement entropy $S(x)$ for the (e) Ising and (f) three-state Potts models with $L=96$ and different deformation strengths $\theta$. Colored markers are numerical results, and solid curves are analytical predictions from Eq.~\eqref{eq:qMobius-EE}.}
\label{fig:ESEE}
\end{figure}

As in the uniform case~\cite{Laeuchli2013}, the entanglement spectrum~\cite{LiH2008} of the deformed ground state retains the universal boundary-CFT (BCFT) tower structure. Denoting the entanglement energies by $\varepsilon_\alpha$ (i.e., eigenvalues of $-\ln \rho_A$), their universal gaps are
\begin{align}
    \varepsilon_\alpha-\varepsilon_0
    = \frac{2\pi^2}{W_A^{(q,\theta)}(x_1,x_2)}
    \left( \Delta_\alpha-\Delta_0 \right),
\label{eq:qMobius-ES-gap}
\end{align}
where $\Delta_{\alpha}$ are conformal dimensions~\cite{SM}. 
Thus, the $q$-M\"obius deformation modifies the entanglement spectrum only through the geometric factor $W_A^{(q,\theta)}(x_1,x_2)$, leaving the universal BCFT level structure unchanged. 
As shown in Fig.~\ref{fig:ESEE}(a-d), the low-lying entanglement spectra of the critical Ising and three-state Potts chains clearly exhibit the predicted BCFT towers.

The same reduced density operator also determines the entanglement entropy. In the annulus representation, $\rho_A$ is a thermal density operator generated by the open-string channel Virasoro zero mode~\cite{SM}. Its inverse temperature is fixed by the annulus width as $\beta_A=2\pi^2/W_A^{(q,\theta)}(x_1,x_2)$. 
Following the definition of von Neumann entropy $S_A=-\tr(\rho_A\ln\rho_A)$, the entanglement entropy is
\begin{align}
 S_A(x_1,x_2)  &= \frac{c}{6}W_A^{(q,\theta)}(x_1,x_2) + s_1,
\label{eq:qMobius-EE}
\end{align}
where $s_1$ is a nonuniversal constant~\cite{SM}. Equation~\eqref{eq:qMobius-EE} shows that the $q$-M\"obius deformation modifies the entanglement only through the endpoint Jacobians and the re-parametrized interval length. As shown in Fig.~\ref{fig:ESEE}(e,f), the numerically computed entanglement entropies of the critical Ising and three-state Potts chains agree well with the analytical prediction in Eq.~\eqref{eq:qMobius-EE}~\cite{EEUVcutoff}. Together with the entanglement spectra shown in Fig.~\ref{fig:ESEE}(a-d), these results indicate that the $q$-M\"obius ground states remain moderately entangled, thereby justifying the use of DMRG for approximating the deformed ground states.

\begin{figure}
    \centering
    \includegraphics[width=1.0\linewidth]{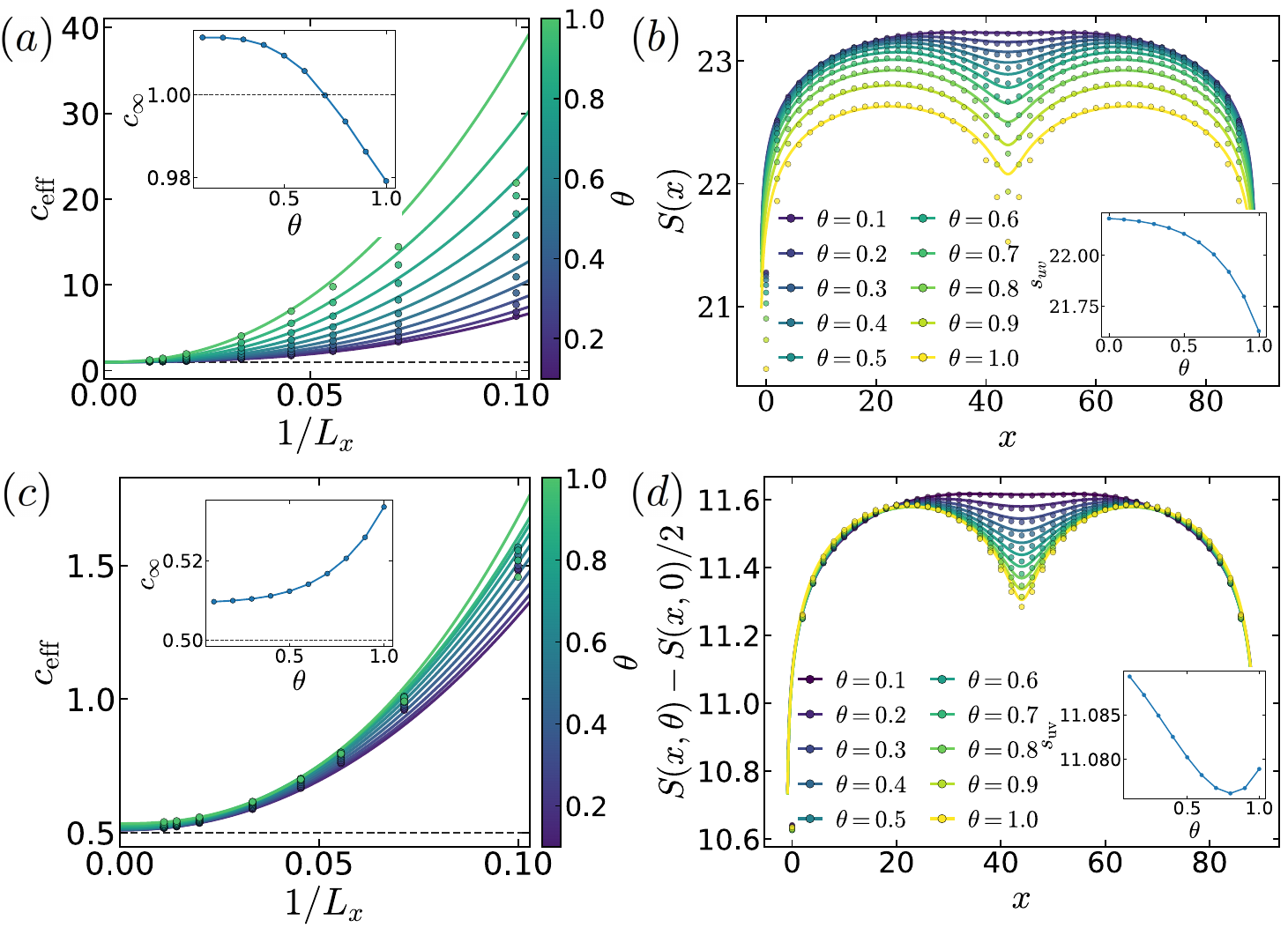}
    \caption{Edge-state central charge and entanglement entropy in the $q=2$ deformed QWZ model.
    (a) Finite-size extrapolation of the effective central charge $c_{\mathrm{eff}}$ for the two-edge deformation, using $L_x=10,14,18,22,30,50,70,90$ at fixed $L_y=30$. 
    Colored markers denote numerical data for different deformation strengths $\theta$, and solid curves show finite-size fits of the form $aL^{-p}+b$ using the four largest system sizes.
    The inset shows the thermodynamic-limit estimate $c_{\infty}(\theta)$, with the dashed line indicating the expected CFT value.
    (b) Entanglement entropy profile $S(x)$ for the two-edge deformation. 
    Colored markers are numerical data, and solid curves are fits to Eq.~\eqref{eq:qMobius-EE}.
    The inset shows the fitted nonuniversal constant as a function of $\theta$.
    (c) Finite-size extrapolation of $c_{\mathrm{eff}}$ for a single-edge deformation, where the deformation is localized near one boundary and decays exponentially into the bulk along the $y$ direction.
    The same extrapolation procedure as in panel (a) is used, and the inset shows $c_{\infty}(\theta)$.
    (d) Entanglement entropy associated with the deformed edge after subtracting the contribution from the undeformed edge. 
    The data are fitted by Eq.~\eqref{eq:qMobius-EE}, with the coefficient corresponding to a single chiral edge mode, and the inset shows the fitted nonuniversal constant.}
    \label{fig:QWZ}
\end{figure}

\textit{Generalization to Higher Dimensions}.
The overlap construction can also be generalized to 2D topological phases with gapless edge modes. As a representative example, we consider the Qi-Wu-Zhang (QWZ) model on the square lattice~\cite{QiXL2006}
\begin{align}
     \tilde{H}_{\mathrm{QWZ}} &= m\sum_{\mathbf{r}}c_{\mathbf{r}}^\dagger \sigma^z c_{\mathbf{r}} + \frac{1}{2} \sum_{\mathbf{r}}\left[c_{\mathbf{r}+\hat{x}}^\dagger (\sigma^z-{\rm i}\sigma^x) c_{\mathbf{r}}\right. \nonumber \\
    &\phantom{=} \; \left.+c_{\mathbf{r}+\hat{y}}^\dagger (\sigma^z-{\rm i}\sigma^y) c_{\mathbf{r}}+ \mathrm{h.c.} \right],
\label{eq:QWZ-Hamiltonian}
\end{align}
where $c_{\mathbf r}=(c_{\mathbf r,\uparrow},c_{\mathbf r,\downarrow})^T$, and $\mathbf r=(x,y)$ labels the lattice sites. 
The Pauli matrices $\sigma^{x,y,z}$ act on the two internal components of $c_{\mathbf r}$. 
For $-2<m<0$, the QWZ model is a Chern insulator with Chern number $C=-1$. 
On a cylinder, it supports one gapless edge mode, a chiral Dirac fermion, on each boundary. 
In the following, we set $m=-1$ and study the QWZ model on a cylinder, with antiperiodic (open) boundary conditions along the $x$ ($y$) direction, choosing even $L_x$ to avoid edge zero modes and place the edge CFT in the Neveu-Schwarz sector. We focus on the quasi-one-dimensional limit with fixed $L_y$ and $L_x\to\infty$. In this limit, the deformation becomes slowly varying on the scale of the finite bulk correlation length, so the gapped bulk is expected to contribute only nonuniversal finite-size corrections, while the overlap is governed by the gapless edge CFT.

To deform the 2D Hamiltonian, we generalize the lattice construction in Eq.~\eqref{eq:lattice-Hq} by applying the same envelope function to local Hamiltonian terms according to their spatial positions. 
In Eq.~\eqref{eq:QWZ-Hamiltonian}, the onsite mass term and the hopping term along the $y$ direction are assigned to lattice sites, while the hopping term along the $x$ direction is assigned to nearest-neighbor bonds. 
Since the QWZ Hamiltonian is quadratic in fermion operators, its ground states are Slater determinants, and their overlaps can be efficiently evaluated using the standard determinant formula~\cite{Freefermion}.
Figure~\ref{fig:QWZ}(a) shows the effective central charge extracted from the deformed QWZ Hamiltonian for different deformation strengths $\theta$ and system sizes $L_x$, with $L_y=30$ fixed. 
The finite-size extrapolation to $L_x\to\infty$ yields values close to $c=1$, as shown in the inset. 
This reflects that the two spatially separated edge modes on the cylinder form the chiral and anti-chiral sectors of an effective $c=1$ CFT.

Using correlation-matrix techniques for free-fermion systems~\cite{Peschel2003}, we further compute the entanglement entropy of the subsystem $[0,x]\times[0,L_y-1]$ at $L_x=90$ for different deformation strengths $\theta$; see Fig.~\ref{fig:QWZ}(b). 
The numerical results agree well with the analytical prediction in Eq.~\eqref{eq:qMobius-EE}, supporting the effective CFT description of the deformed edge states. 
Unlike the purely 1D examples above, the fitted nonuniversal constant $s_{\mathrm{uv}}$ shown in the inset exhibits a weak $\theta$ dependence, likely due to residual contributions from gapped bulk degrees of freedom.

A key difference from 1D critical chains is that the two chiral sectors of the gapless edge theory are spatially separated in the 2D Chern insulator. 
This allows us to selectively deform only one edge and probe a single chiral sector. 
To implement this, we multiply the modulated part of the envelope function by an exponential factor $e^{-y/\xi_y}$ with $\xi_y=10$, so that the deformation is localized near the edge at $y=0$ and becomes negligible for $y\gg \xi_y$.

The corresponding results are shown in Fig.~\ref{fig:QWZ}(c,d). 
Figure~\ref{fig:QWZ}(c) shows that extrapolation of the effective central charge extracted from Eq.~\eqref{eq:Central-Charge-Lattice} yields values close to $c=1/2$. 
Here we use the convention $c=(c_L+c_R)/2$, while the chiral central charge is $c_-=c_L-c_R$. 
For a single QWZ edge, only one chiral sector contributes; taking $c_R=0$, this result corresponds to $c_-=2c=1$.
In Fig.~\ref{fig:QWZ}(d), we plot the entanglement entropy associated with the deformed edge after subtracting the contribution from the undeformed edge. 
The resulting entropy profile is well described by Eq.~\eqref{eq:qMobius-EE}, with the coefficient corresponding to a single chiral edge mode with $c_-=1$. 
The fitted nonuniversal term, shown in the inset, remains nearly independent of the deformation strength $\theta$.

\textit{Summary and Outlook}.
To summarize, we have introduced a $q$-M\"obius ground-state-overlap estimator for the central charge whose universal CFT prediction is independent of system size and nonuniversal velocity. Its lattice realization yields accurate estimates from a single sufficiently large system for representative critical chains and extends to the gapless edges of a 2D Chern insulator.
The entanglement spectrum and entanglement entropy further confirm that the deformed states retain the predicted conformal geometry while remaining logarithmically entangled and numerically tractable within DMRG.

Looking ahead, controlling finite-size corrections will be important for identifying the optimal deformation window and improving numerical precision. 
Our results for scaling dimensions and gapless edges motivate exploring broader applications of spatial deformations as wave-function probes of conformal data, particularly in 2D topological phases.
The unitary origin of the deformation also points to a possible experimental direction: an approximate lattice implementation of $U_q(\theta)$ may provide a circuit-based route to the deformed ground state, while spatially modulated Hamiltonians may be realizable in programmable quantum simulators~\cite{Daley2022}. 
If the uniform and deformed ground states can be prepared with sufficient fidelity, the overlap entering our estimator could in principle be accessed using interferometric protocols~\cite{Ekert2002} or randomized-measurement protocols~\cite{Tiff2019,Elben2023}.

\textit{Data availability}. The data that support the findings
of this article are openly available at~\cite{BaiC-Data}.

\textit{Acknowledgments}. We thank Bastien Lapierre, Zehan Li, Weibo Mao, Masahiro Nozaki, Jia Tian, Yueshui Zhang, Yuxuan Zhang, Zihan Zhou for illuminating discussions. We thank Jan von Delft and Meng Cheng for their valuable feedback on the manuscript. C.B. acknowledges financial support from Masahiro Nozaki at the University of Chinese Academy of Sciences, and thanks Shanxi University for its hospitality during the preparation of this work.
Numerical tensor-network calculations were performed using the TeNPy Library~\cite{tenpy2024}.

\bibliography{refs.bib}

@article{Miyata2024,
    author = "Miyata, Akihiro and Nozaki, Masahiro and Tamaoka, Kotaro and Watanabe, Masataka",
    title = "{Hawking-Page and entanglement phase transition in 2d CFT on curved backgrounds}",
    reportNumber = "RIKEN-iTHEMS-Report-24",
    doi = "10.1007/JHEP08(2024)190",
    journal = "J. High Energy Phys.",
    volume = "08",
    pages = "190",
    year = "2024"
}

@Article{White1992,
  Title                    = {Density matrix formulation for quantum renormalization groups},
  Author                   = {White, Steven R.},
  Journal                  = {Phys. Rev. Lett.},
  Year                     = {1992},
  Month                    = {Nov},
  Pages                    = {2863--2866},
  Volume                   = {69},
  Doi                      = {10.1103/PhysRevLett.69.2863},
  Issue                    = {19},
  Numpages                 = {0},
  Publisher                = {American Physical Society},
  Url                      = {http://link.aps.org/doi/10.1103/PhysRevLett.69.2863}
}

@Article{Schollwoeck2011,
  Title                    = {The density-matrix renormalization group in the age of matrix product states},
  Author                   = {Ulrich Schollw{\"o}ck},
  Journal                  = {Ann. Phys.},
  Year                     = {2011},
  Number                   = {1},
  Pages                    = {96 - 192},
  Volume                   = {326},
  Doi                      = {10.1016/j.aop.2010.09.012},
  Url                      = {www.sciencedirect.com/science/article/pii/S0003491610001752}
}

@article{SM,
  journal  = {See Supplemental Material for derivations of the q-M\"obius-deformed ground state, the exact overlap formula, the entanglement spectrum and entropy, and additional numerical results, which includes Refs.~\cite{Affleck1991,Blume1966,Capel1966,Xavier2012,Albertini1992,Gomez1996,Dotsenko1984,Kedem1993,LiW2015,Affleck1988,Fuehringer2008,Aguado2009}}
}

@dataset{BaiC-Data,
  author       = {Bai, Chen and
                  Sun, Xinyu and
                  Mo, Liang-Hong and
                  Tu, Hong-Hao},
  title        = {Extracting central charge from ground-state overlaps of spatially deformed Hamiltonians [Data set]
                  },
  month        = {June},
  year         = {2026},
  publisher    = {Zenodo},
  doi          = {10.5281/zenodo.20580660},
  url          = {https://doi.org/10.5281/zenodo.20580660},
}

@article{Holzhey1994,
    author = "Holzhey, Christoph and Larsen, Finn and Wilczek, Frank",
    title = "{Geometric and renormalized entropy in conformal field theory}",
    doi = "10.1016/0550-3213(94)90402-2",
    journal = "Nucl. Phys. B",
    volume = "424",
    pages = "443--467",
    year = "1994"
}

@article{Kobayashi2024,
  title = {Extracting Higher Central Charge from a Single Wave Function},
  author = {Kobayashi, Ryohei and Wang, Taige and Soejima, Tomohiro and Mong, Roger S. K. and Ryu, Shinsei},
  journal = {Phys. Rev. Lett.},
  volume = {132},
  issue = {1},
  pages = {016602},
  numpages = {6},
  year = {2024},
  month = {Jan},
  publisher = {American Physical Society},
  doi = {10.1103/PhysRevLett.132.016602},
  url = {https://link.aps.org/doi/10.1103/PhysRevLett.132.016602}
}

@article{KimIH2022a,
  title = {Chiral Central Charge from a Single Bulk Wave Function},
  author = {Kim, Isaac H. and Shi, Bowen and Kato, Kohtaro and Albert, Victor V.},
  journal = {Phys. Rev. Lett.},
  volume = {128},
  issue = {17},
  pages = {176402},
  numpages = {6},
  year = {2022},
  month = {Apr},
  publisher = {American Physical Society},
  doi = {10.1103/PhysRevLett.128.176402},
  url = {https://link.aps.org/doi/10.1103/PhysRevLett.128.176402}
}

@article{KimIH2022b,
  title = {Modular commutator in gapped quantum many-body systems},
  author = {Kim, Isaac H. and Shi, Bowen and Kato, Kohtaro and Albert, Victor V.},
  journal = {Phys. Rev. B},
  volume = {106},
  issue = {7},
  pages = {075147},
  numpages = {17},
  year = {2022},
  month = {Aug},
  publisher = {American Physical Society},
  doi = {10.1103/PhysRevB.106.075147},
  url = {https://link.aps.org/doi/10.1103/PhysRevB.106.075147}
}

@article{FanRH2022,
  title = {From Entanglement Generated Dynamics to the Gravitational Anomaly and Chiral Central Charge},
  author = {Fan, Ruihua},
  journal = {Phys. Rev. Lett.},
  volume = {129},
  issue = {26},
  pages = {260403},
  numpages = {6},
  year = {2022},
  month = {Dec},
  publisher = {American Physical Society},
  doi = {10.1103/PhysRevLett.129.260403},
  url = {https://link.aps.org/doi/10.1103/PhysRevLett.129.260403}
}

@article{ZouYJ2022a,
  title = {Universal information of critical quantum spin chains from wavefunction overlap},
  author = {Zou, Yijian},
  journal = {Phys. Rev. B},
  volume = {105},
  issue = {16},
  pages = {165420},
  numpages = {11},
  year = {2022},
  month = {Apr},
  publisher = {American Physical Society},
  doi = {10.1103/PhysRevB.105.165420},
  url = {https://link.aps.org/doi/10.1103/PhysRevB.105.165420}
}

@article{ZouYJ2022b,
  title = {Modular Commutators in Conformal Field Theory},
  author = {Zou, Yijian and Shi, Bowen and Sorce, Jonathan and Lim, Ian T. and Kim, Isaac H.},
  journal = {Phys. Rev. Lett.},
  volume = {129},
  issue = {26},
  pages = {260402},
  numpages = {7},
  year = {2022},
  month = {Dec},
  publisher = {American Physical Society},
  doi = {10.1103/PhysRevLett.129.260402},
  url = {https://link.aps.org/doi/10.1103/PhysRevLett.129.260402}
}

@article{LiuYH2023,
  title = {Operator fusion from wave-function overlap: Universal finite-size corrections and application to the Haagerup model},
  author = {Liu, Yuhan and Zou, Yijian and Ryu, Shinsei},
  journal = {Phys. Rev. B},
  volume = {107},
  issue = {15},
  pages = {155124},
  numpages = {20},
  year = {2023},
  month = {Apr},
  publisher = {American Physical Society},
  doi = {10.1103/PhysRevB.107.155124},
  url = {https://link.aps.org/doi/10.1103/PhysRevB.107.155124}
}

@article{HanB2020,
  title = {Classification of $S{L}_{2}$ deformed Floquet conformal field theories},
  author = {Han, Bo and Wen, Xueda},
  journal = {Phys. Rev. B},
  volume = {102},
  issue = {20},
  pages = {205125},
  numpages = {15},
  year = {2020},
  month = {Nov},
  publisher = {American Physical Society},
  doi = {10.1103/PhysRevB.102.205125},
  url = {https://link.aps.org/doi/10.1103/PhysRevB.102.205125}
}

@article{Katsura2011,
doi = {10.1088/1751-8113/44/25/252001},
url = {https://doi.org/10.1088/1751-8113/44/25/252001},
year = {2011},
month = {may},
publisher = {},
volume = {44},
number = {25},
pages = {252001},
author = {Katsura, Hosho},
title = {Exact ground state of the sine-square deformed XY spin chain},
journal = {J. Phys. A},}

@article{Katsura2012,
    author = "Katsura, Hosho",
    title = "{Sine-square deformation of solvable spin chains and conformal field theories}",
    doi = "10.1088/1751-8113/45/11/115003",
    journal = "J. Phys. A",
    volume = "45",
    pages = "115003",
    year = "2012"
}

@article{Tada2015,
author = {Tada, Tsukasa},
title = {Sine-square deformation and its relevance to string theory},
journal = {Mod. Phys. Lett. A},
volume = {30},
number = {19},
pages = {1550092},
year = {2015},
doi = {10.1142/s0217732315500923},
}

@article{Lapierre2025a,
  title = {Driven nonunitary dynamics of quantum critical systems},
  author = {Lapierre, Bastien and Pelliconi, Pietro and Ryu, Shinsei and Sonner, Julian},
  journal = {Phys. Rev. B},
  volume = {112},
  issue = {10},
  pages = {104322},
  numpages = {18},
  year = {2025},
  month = {Sep},
  publisher = {American Physical Society},
  doi = {10.1103/lwrz-jxrr},
  url = {https://link.aps.org/doi/10.1103/lwrz-jxrr}
}

@article{Lapierre2025b,
    author = "Lapierre, Bastien and Moosavi, Per and Oblak, Blagoje",
    title = "{Nonequilibrium Probes of Quantum Geometry in Gapless Systems}",
    eprint = "2511.09639",
    archivePrefix = "arXiv",
    month = "11",
    year = "2025",
    journal = "",
}

@article{Cardy2014,
  title = {Thermalization and Revivals after a Quantum Quench in Conformal Field Theory},
  author = {Cardy, John},
  journal = {Phys. Rev. Lett.},
  volume = {112},
  issue = {22},
  pages = {220401},
  numpages = {5},
  year = {2014},
  month = {Jun},
  publisher = {American Physical Society},
  doi = {10.1103/PhysRevLett.112.220401},
  url = {https://link.aps.org/doi/10.1103/PhysRevLett.112.220401}
}

@article{Zamolodchikov1986,
  title={``Irreversibility'' of the Flux of the Renormalization Group in a 2D Field Theory},
  author={Zamolodchikov, Alexander B},
  journal={JETP Lett.},
  volume={43},
  number={12},
  pages={730--732},
  year={1986},
  url={http://jetpletters.ru/ps/1413/article_21504.shtml},
}

@article{Vidal2003,
  title = {Entanglement in Quantum Critical Phenomena},
  author = {Vidal, G. and Latorre, J. I. and Rico, E. and Kitaev, A.},
  journal = {Phys. Rev. Lett.},
  volume = {90},
  issue = {22},
  pages = {227902},
  numpages = {4},
  year = {2003},
  month = {Jun},
  publisher = {American Physical Society},
  doi = {10.1103/PhysRevLett.90.227902},
  url = {https://link.aps.org/doi/10.1103/PhysRevLett.90.227902}
}

@article{Calabrese2004,
    author = "Calabrese, Pasquale and Cardy, John L.",
    title = "{Entanglement entropy and quantum field theory}",
    doi = "10.1088/1742-5468/2004/06/P06002",
    journal = "J. Stat. Mech.",
    volume = "0406",
    pages = "P06002",
    year = "2004"
}

@article{Belavin1984,
    author = "Belavin, A. A. and Polyakov, Alexander M. and Zamolodchikov, A. B.",
    editor = "Khalatnikov, I. M. and Mineev, V. P.",
    title = "{Infinite Conformal Symmetry in Two-Dimensional Quantum Field Theory}",
    reportNumber = "CERN-TH-3827",
    doi = "10.1016/0550-3213(84)90052-X",
    journal = "Nucl. Phys. B",
    volume = "241",
    pages = "333--380",
    year = "1984"
}

@article{Bloete1986,
  title = {Conformal invariance, the central charge, and universal finite-size amplitudes at criticality},
  author = {Bl\"ote, H. W. J. and Cardy, John L. and Nightingale, M. P.},
  journal = {Phys. Rev. Lett.},
  volume = {56},
  issue = {7},
  pages = {742--745},
  numpages = {0},
  year = {1986},
  month = {Feb},
  publisher = {American Physical Society},
  doi = {10.1103/PhysRevLett.56.742},
  url = {https://link.aps.org/doi/10.1103/PhysRevLett.56.742}
}

@article{Koo1993,
    author = "Koo, W. M. and Saleur, H.",
    title = "{Representations of the Virasoro algebra from lattice models}",
    reportNumber = "USC-93-025, YCTP-P22-93",
    doi = "10.1016/0550-3213(94)90018-3",
    journal = "Nucl. Phys. B",
    volume = "426",
    pages = "459--504",
    year = "1994"
}

@article{Cardy1984a,
doi = {10.1088/0305-4470/17/7/003},
url = {https://doi.org/10.1088/0305-4470/17/7/003},
year = {1984},
month = {may},
publisher = {},
volume = {17},
number = {7},
pages = {L385},
author = {Cardy, John L.},
title = {Conformal invariance and universality in finite-size scaling},
journal = {J. Phys. A},
}

@article{Milsted2017,
  title = {Extraction of conformal data in critical quantum spin chains using the Koo-Saleur formula},
  author = {Milsted, Ashley and Vidal, Guifre},
  journal = {Phys. Rev. B},
  volume = {96},
  issue = {24},
  pages = {245105},
  numpages = {13},
  year = {2017},
  month = {Dec},
  publisher = {American Physical Society},
  doi = {10.1103/PhysRevB.96.245105},
  url = {https://link.aps.org/doi/10.1103/PhysRevB.96.245105}
}

@article{FanRH2019,
  title = {Emergent Spatial Structure and Entanglement Localization in Floquet Conformal Field Theory},
  author = {Fan, Ruihua and Gu, Yingfei and Vishwanath, Ashvin and Wen, Xueda},
  journal = {Phys. Rev. X},
  volume = {10},
  issue = {3},
  pages = {031036},
  numpages = {25},
  year = {2020},
  month = {Aug},
  publisher = {American Physical Society},
  doi = {10.1103/PhysRevX.10.031036},
  url = {https://link.aps.org/doi/10.1103/PhysRevX.10.031036}
}

@article{Hikihara2011,
  title = {Connecting distant ends of one-dimensional critical systems by a sine-square deformation},
  author = {Hikihara, Toshiya and Nishino, Tomotoshi},
  journal = {Phys. Rev. B},
  volume = {83},
  issue = {6},
  pages = {060414(R)},
  numpages = {4},
  year = {2011},
  month = {Feb},
  publisher = {American Physical Society},
  doi = {10.1103/PhysRevB.83.060414},
  url = {https://link.aps.org/doi/10.1103/PhysRevB.83.060414}
}

@article{Maruyama2011,
  title = {Sine-square deformation of free fermion systems in one and higher dimensions},
  author = {Maruyama, Isao and Katsura, Hosho and Hikihara, Toshiya},
  journal = {Phys. Rev. B},
  volume = {84},
  issue = {16},
  pages = {165132},
  numpages = {8},
  year = {2011},
  month = {Oct},
  publisher = {American Physical Society},
  doi = {10.1103/PhysRevB.84.165132},
  url = {https://link.aps.org/doi/10.1103/PhysRevB.84.165132}
}

@article{Okunishi2016,
    author = "Okunishi, Kouichi",
    title = {{Sine-square deformation and M{\"o}bius quantization of 2D conformal field theory}},
    doi = "10.1093/ptep/ptw060",
    journal = "Prog. Theor. Exp. Phys.",
    volume = "2016",
    number = "6",
    pages = "063A02",
    year = "2016"
}

@book{DiFrancesco1997,
  author    = {P. Di Francesco and P. Mathieu and D. S\'en\'echal},
  publisher = {Springer-Verlag, New York},
  title     = {Conformal Field Theory},
  year      = {1997},
  doi       = {10.1007/978-1-4612-2256-9},
}

@Article{BaiC2024,
  author    = {Bai, Chen and Miyata, Akihiro and Nozaki, Masahiro},
  journal   = {J. High Energy Phys.},
  title     = {Entanglement dynamics in 2d HCFTs on the curved background: the case of q-M{\"o}bius Hamiltonian},
  year      = {2024},
  month     = {Dec},
  pages     = {208},
  volume    = {12},
  doi       = {10.1007/JHEP12(2024)208},
  publisher = {Springer},
  url       = {https://link.springer.com/article/10.1007/JHEP12(2024)208},
}

@article{Tamura2017,
    author = "Tamura, Shota and Katsura, Hosho",
    title = "{Zero-energy states in conformal field theory with sine-square deformation}",
    doi = "10.1093/ptep/ptx147",
    journal = "Prog. Theor. Exp. Phys.",
    volume = "2017",
    number = "11",
    pages = "113A01",
    year = "2017"
}

@article{TuHH2013,
  title = {Momentum polarization: An entanglement measure of topological spin and chiral central charge},
  author = {Tu, Hong-Hao and Zhang, Yi and Qi, Xiao-Liang},
  journal = {Phys. Rev. B},
  volume = {88},
  issue = {19},
  pages = {195412},
  numpages = {12},
  year = {2013},
  month = {Nov},
  publisher = {American Physical Society},
  doi = {10.1103/PhysRevB.88.195412},
  url = {https://link.aps.org/doi/10.1103/PhysRevB.88.195412}
}

@article{Zaletel2013,
  title = {Topological Characterization of Fractional Quantum Hall Ground States from Microscopic Hamiltonians},
  author = {Zaletel, Michael P. and Mong, Roger S. K. and Pollmann, Frank},
  journal = {Phys. Rev. Lett.},
  volume = {110},
  issue = {23},
  pages = {236801},
  numpages = {5},
  year = {2013},
  month = {Jun},
  publisher = {American Physical Society},
  doi = {10.1103/PhysRevLett.110.236801},
  url = {https://link.aps.org/doi/10.1103/PhysRevLett.110.236801}
}

@article{QiXL2006,
  title = {Topological quantization of the spin Hall effect in two-dimensional paramagnetic semiconductors},
  author = {Qi, Xiao-Liang and Wu, Yong-Shi and Zhang, Shou-Cheng},
  journal = {Phys. Rev. B},
  volume = {74},
  issue = {8},
  pages = {085308},
  numpages = {7},
  year = {2006},
  month = {Aug},
  publisher = {American Physical Society},
  doi = {10.1103/PhysRevB.74.085308},
  url = {https://link.aps.org/doi/10.1103/PhysRevB.74.085308}
}

@Article{Uimin1970,
  author  = {Uimin, G. V.},
  journal = {JETP Lett.},
  title   = {One-dimensional Problem for S = 1 with Modified Antiferromagnetic Hamiltonian},
  year    = {1970},
  pages   = {332},
  volume  = {12},
  issue   = {6},
  url     = {http://jetpletters.ru/ps/1730/article_26296.shtml},
}

@article{LaiCK1974,
author = {Lai, C. K.},
title = {Lattice gas with nearest‐neighbor interaction in one dimension with arbitrary statistics},
journal = {J. Math. Phys.},
volume = {15},
number = {10},
pages = {1675-1676},
year = {1974},
doi = {10.1063/1.1666522},
URL = {https://doi.org/10.1063/1.1666522}
}

@article{Sutherland1975,
  title = {Model for a multicomponent quantum system},
  author = {Sutherland, Bill},
  journal = {Phys. Rev. B},
  volume = {12},
  issue = {9},
  pages = {3795--3805},
  numpages = {0},
  year = {1975},
  month = {Nov},
  publisher = {American Physical Society},
  doi = {10.1103/PhysRevB.12.3795},
  url = {https://link.aps.org/doi/10.1103/PhysRevB.12.3795}
}

@article{LiH2008,
  title = {Entanglement Spectrum as a Generalization of Entanglement Entropy: Identification of Topological Order in Non-Abelian Fractional Quantum Hall Effect States},
  author = {Li, Hui and Haldane, F. D. M.},
  journal = {Phys. Rev. Lett.},
  volume = {101},
  issue = {1},
  pages = {010504},
  numpages = {4},
  year = {2008},
  month = {Jul},
  publisher = {American Physical Society},
  doi = {10.1103/PhysRevLett.101.010504},
  url = {https://link.aps.org/doi/10.1103/PhysRevLett.101.010504}
}

@article{Laeuchli2013,
      title={Operator content of real-space entanglement spectra at conformal critical points}, 
      author={Andreas M. L\"auchli},
      year={2013},
      eprint={1303.0741},
      archivePrefix={arXiv},
      journal={ }, 
}

@article{Peschel2003,
doi = {10.1088/0305-4470/36/14/101},
url = {https://doi.org/10.1088/0305-4470/36/14/101},
year = {2003},
month = {mar},
publisher = {},
volume = {36},
number = {14},
pages = {L205},
author = {Ingo Peschel},
title = {Calculation of reduced density matrices from correlation functions},
journal = {J. Phys. A},
}

@article{MoLH2026,
      title={Observing conformal Floquet dynamics on a digital quantum processor}, 
      author={Liang-Hong Mo and Bastien Lapierre and Qiang Miao},
      year={2026},
      eprint={2605.27530},
      archivePrefix={arXiv},
      journal={ }
}

@article{UVcutoff,
  journal  = {The spectral gaps are fitted using the analytical expression with a non-universal UV constant, which reflects the dependence on the UV cutoff $\epsilon$ in $W_A(q,\theta)$}
}

@article{EEUVcutoff,
  journal  = {The entanglement entropy for different deformation strengths $\theta$ is fitted using a common UV constant, suggesting that this UV contribution is invariant under the deformation}
}

@article{Freefermion,
  journal  = {For two Slater-determinant states $|\psi_1\rangle = \prod_{m=1}^{M} d_{m}^{\dagger}|0\rangle$ and $|\psi_2\rangle = \prod_{m=1}^{M} f_{m}^{\dagger}|0\rangle$, with $d^{\dagger}_m = \sum_i c^{\dagger}_i (V_1)_{i,m}$ and $f^{\dagger}_m = \sum_i c^{\dagger}_i (V_2)_{i,m}$, the overlap is given by $\langle \psi_1 | \psi_2 \rangle = \mathrm{det}(V^{\dagger}_1 V_2)$}
}

@Article{tenpy2024,
    title={{Tensor network Python (TeNPy) version 1}},
    author={Johannes Hauschild and Jakob Unfried and Sajant Anand and Bartholomew Andrews and Marcus Bintz and Umberto Borla and Stefan Divic and Markus Drescher and Jan Geiger and Martin Hefel and Kévin Hémery and Wilhelm Kadow and Jack Kemp and Nico Kirchner and Vincent S. Liu and Gunnar Möller and Daniel Parker and Michael Rader and Anton Romen and Samuel Scalet and Leon Schoonderwoerd and Maximilian Schulz and Tomohiro Soejima and Philipp Thoma and Yantao Wu and Philip Zechmann and Ludwig Zweng and Roger S. K. Mong and Michael P. Zaletel and Frank Pollmann},
    journal={SciPost Phys. Codebases},
    pages={41},
    year={2024},
    publisher={SciPost},
    doi={10.21468/SciPostPhysCodeb.41},
    url={https://scipost.org/10.21468/SciPostPhysCodeb.41},
}

@article{Xavier2012,
  title = {Finite-size corrections of the entanglement entropy of critical quantum chains},
  author = {Xavier, J. C. and Alcaraz, F. C.},
  journal = {Phys. Rev. B},
  volume = {85},
  issue = {2},
  pages = {024418},
  numpages = {8},
  year = {2012},
  month = {Jan},
  publisher = {American Physical Society},
  doi = {10.1103/PhysRevB.85.024418},
  url = {https://link.aps.org/doi/10.1103/PhysRevB.85.024418}
}

@article{Blume1966,
  title = {Theory of the First-Order Magnetic Phase Change in U${\mathrm{O}}_{2}$},
  author = {Blume, M.},
  journal = {Phys. Rev.},
  volume = {141},
  issue = {2},
  pages = {517--524},
  numpages = {0},
  year = {1966},
  month = {Jan},
  publisher = {American Physical Society},
  doi = {10.1103/PhysRev.141.517},
  url = {https://link.aps.org/doi/10.1103/PhysRev.141.517}
}

@article{Capel1966,
title = {On the possibility of first-order phase transitions in Ising systems of triplet ions with zero-field splitting},
journal = {Physica},
volume = {32},
number = {5},
pages = {966-988},
year = {1966},
issn = {0031-8914},
doi = {https://doi.org/10.1016/0031-8914(66)90027-9},
url = {https://www.sciencedirect.com/science/article/pii/0031891466900279},
author = {H.W. Capel},
}

@article{Fuehringer2008,
  title={DMRG studies of critical SU (N) spin chains},
  author={F{\"u}hringer, Max and Rachel, Stephan and Thomale, Ronny and Greiter, Martin and Schmitteckert, Peter},
  journal={Ann. Phys. (Berlin)},
  volume={520},
  number={12},
  pages={922--936},
  year={2008},
  publisher={Wiley Online Library},
  Url={ https://doi.org/10.1002/andp.20085201204}
}

@article{Aguado2009,
  title = {Density-matrix renormalization-group simulation of the $\mathit{SU}(3)$ antiferromagnetic Heisenberg model},
  author = {Aguado, M. and Asorey, M. and Ercolessi, E. and Ortolani, F. and Pasini, S.},
  journal = {Phys. Rev. B},
  volume = {79},
  issue = {1},
  pages = {012408},
  numpages = {4},
  year = {2009},
  month = {Jan},
  publisher = {American Physical Society},
  doi = {10.1103/PhysRevB.79.012408},
  url = {https://link.aps.org/doi/10.1103/PhysRevB.79.012408}
}

@article{Affleck1986,
  title = {Universal term in the free energy at a critical point and the conformal anomaly},
  author = {Affleck, Ian},
  journal = {Phys. Rev. Lett.},
  volume = {56},
  issue = {7},
  pages = {746--748},
  numpages = {0},
  year = {1986},
  month = {Feb},
  publisher = {American Physical Society},
  doi = {10.1103/PhysRevLett.56.746},
  url = {https://link.aps.org/doi/10.1103/PhysRevLett.56.746}
}

@article{Affleck1988,
title = {Critical behaviour of SU(n) quantum chains and topological non-linear $\sigma$-models},
author = {Affleck, Ian},
journal = {Nucl. Phys. B},
volume = {305},
number = {4},
pages = {582-596},
year = {1988},
issn = {0550-3213},
doi = {10.1016/0550-3213(88)90117-4},
url = {https://www.sciencedirect.com/science/article/pii/0550321388901174}
}

@article{Affleck1991,
  title = {Universal noninteger ``ground-state degeneracy'' in critical quantum systems},
  author = {Affleck, Ian and Ludwig, Andreas W. W.},
  journal = {Phys. Rev. Lett.},
  volume = {67},
  issue = {2},
  pages = {161--164},
  numpages = {0},
  year = {1991},
  month = {Jul},
  publisher = {American Physical Society},
  doi = {10.1103/PhysRevLett.67.161},
  url = {https://link.aps.org/doi/10.1103/PhysRevLett.67.161}
}

@book{Gomez1996,
    author = "G\'omez, C. and Ruiz-Altaba, M. and Sierra, G.",
    title = "{Quantum groups in two-dimensional physics}",
    doi = "10.1017/CBO9780511628825",
    publisher = "Cambridge University Press",
    year = "1996"
}

@article{Albertini1992,
title = {Spectrum doubling and the extended Brillouin zone in the excitations of the three state Potts spin chain},
journal = {Phys. Lett. A},
volume = {170},
number = {5},
pages = {397-403},
year = {1992},
issn = {0375-9601},
doi = {10.1016/0375-9601(92)90894-R},
author = {Giuseppe Albertini and Srinandan Dasmahapatra and Barry M. McCoy},
}

@article{Dotsenko1984,
title = {Conformal algebra and multipoint correlation functions in 2D statistical models},
journal = {Nucl. Phys. B},
volume = {240},
number = {3},
pages = {312-348},
year = {1984},
issn = {0550-3213},
doi = {https://doi.org/10.1016/0550-3213(84)90269-4},
url = {https://www.sciencedirect.com/science/article/pii/0550321384902694},
author = {Vl.S. Dotsenko and V.A. Fateev},
}

@article{Kedem1993,
  title={Construction of modular branching functions from Bethe's equations in the 3-state Potts chain},
  author={Kedem, Rinat and McCoy, Barry M},
  journal={J. Stat. Phys.},
  volume={71},
  number={5},
  pages={865--901},
  year={1993},
  publisher={Springer},
  doi={10.1007/BF01049953}
}

@article{LiW2015,
  title = {Criticality in translation-invariant parafermion chains},
  author = {Li, Wei and Yang, Shuo and Tu, Hong-Hao and Cheng, Meng},
  journal = {Phys. Rev. B},
  volume = {91},
  issue = {11},
  pages = {115133},
  numpages = {9},
  year = {2015},
  month = {Mar},
  publisher = {American Physical Society},
  doi = {10.1103/PhysRevB.91.115133},
  url = {https://link.aps.org/doi/10.1103/PhysRevB.91.115133}
}

@article{Daley2022,
  title={Practical quantum advantage in quantum simulation},
  author={Daley, Andrew J and Bloch, Immanuel and Kokail, Christian and Flannigan, Stuart and Pearson, Natalie and Troyer, Matthias and Zoller, Peter},
  journal={Nature},
  volume={607},
  number={7920},
  pages={667--676},
  year={2022},
  publisher={Nature Publishing Group UK London},
  url = {https://www.nature.com/articles/s41586-022-04940-6}
}

@article{Ekert2002,
  title = {Direct Estimations of Linear and Nonlinear Functionals of a Quantum State},
  author = {Ekert, Artur K. and Alves, Carolina Moura and Oi, Daniel K. L. and Horodecki, Micha\l{} and Horodecki, Pawe\l{} and Kwek, L. C.},
  journal = {Phys. Rev. Lett.},
  volume = {88},
  issue = {21},
  pages = {217901},
  numpages = {4},
  year = {2002},
  month = {May},
  publisher = {American Physical Society},
  doi = {10.1103/PhysRevLett.88.217901},
  url = {https://link.aps.org/doi/10.1103/PhysRevLett.88.217901}
}

@article{Tiff2019,
author = {Tiff Brydges  and Andreas Elben  and Petar Jurcevic  and Benoît Vermersch  and Christine Maier  and Ben P. Lanyon  and Peter Zoller  and Rainer Blatt  and Christian F. Roos },
title = {Probing Rényi entanglement entropy via randomized measurements},
journal = {Science},
volume = {364},
number = {6437},
pages = {260-263},
year = {2019},
doi = {10.1126/science.aau4963},
}

@article{Elben2023,
  title={The randomized measurement toolbox},
  author={Elben, Andreas and Flammia, Steven T and Huang, Hsin-Yuan and Kueng, Richard and Preskill, John and Vermersch, Beno{\^\i}t and Zoller, Peter},
  journal={Nat. Rev. Phys.},
  volume={5},
  number={1},
  pages={9--24},
  year={2023},
  publisher={Nature Publishing Group UK London},
  URL = {https://www.nature.com/articles/s42254-022-00535-2},
}

@article{CiracSierra2010,
  title = {Infinite matrix product states, conformal field theory, and the {Haldane-Shastry} model},
  author = {Cirac, J. Ignacio and Sierra, Germ{\'a}n},
  journal = {Phys. Rev. B},
  volume = {81},
  number = {10},
  pages = {104431},
  year = {2010},
  month = {Mar},
  publisher = {American Physical Society},
  doi = {10.1103/PhysRevB.81.104431},
  url = {https://doi.org/10.1103/PhysRevB.81.104431},
}

@article{NielsenCiracSierra2011,
  title = {Quantum spin Hamiltonians for the {$\mathrm{SU}(2)_k$} {WZW} model},
  author = {Nielsen, Anne E. B. and Cirac, J. Ignacio and Sierra, Germ{\'a}n},
  journal = {J. Stat. Mech.: Theory Exp.},
  volume = {2011},
  number = {11},
  pages = {P11014},
  year = {2011},
  month = {Nov},
  doi = {10.1088/1742-5468/2011/11/P11014},
  url = {https://doi.org/10.1088/1742-5468/2011/11/P11014},
}

@Article{Haldane1988a,
  author    = {Haldane, F. D. M.},
  journal   = {Phys. Rev. Lett.},
  title     = {Exact Jastrow-Gutzwiller resonating-valence-bond ground state of the spin-1/2 antiferromagnetic Heisenberg chain with 1/$r^{2}$ exchange},
  year      = {1988},
  month     = {Feb},
  pages     = {635--638},
  volume    = {60},
  doi       = {10.1103/PhysRevLett.60.635},
  issue     = {7},
  numpages  = {0},
  publisher = {American Physical Society},
  url       = {https://link.aps.org/doi/10.1103/PhysRevLett.60.635},
}

@article{Shastry1988,
  title = {Exact solution of an S=1/2 Heisenberg antiferromagnetic chain with long-ranged interactions},
  author = {Shastry, B. Sriram},
  journal = {Phys. Rev. Lett.},
  volume = {60},
  issue = {7},
  pages = {639--642},
  numpages = {0},
  year = {1988},
  month = {Feb},
  publisher = {American Physical Society},
  doi = {10.1103/PhysRevLett.60.639},
  url = {https://link.aps.org/doi/10.1103/PhysRevLett.60.639}
}

\clearpage

\appendix

\begin{widetext}

\begin{center}
\textbf{Supplemental Material for ``Extracting central charge from ground-state overlaps of spatially deformed Hamiltonians''}
\end{center}

\setcounter{table}{0}
\renewcommand{\thetable}{S\arabic{table}}
\setcounter{figure}{0}
\renewcommand{\thefigure}{S\arabic{figure}}
\setcounter{equation}{0}
\renewcommand{\theequation}{S\arabic{equation}}

This Supplemental Material provides technical details supporting the main text. 
In Sec.~I, we derive the $q$-M\"obius Hamiltonian from the Virasoro algebra and obtain the exact CFT ground-state overlap. In Sec.~II, we derive the reduced density operator of the $q$-M\"obius ground state by mapping the slit cylinder to an annulus, and obtain the corresponding entanglement Hamiltonian, entanglement spectrum, normalized entanglement gaps, and von Neumann entanglement entropy.
In Sec.~III, we present additional numerical results supporting the conclusions of the main text. Section~IV develops the $q$-M\"obius deformation of general infinite-dimensional matrix product state wave functions, and then specializes the construction to the spin-1/2 Haldane-Shastry state, proving the deformed state to be an exact ground state of an explicit nonlocal parent Hamiltonian.

\tableofcontents

\section{I. CFT derivation of wave-function overlaps}
\label{sec:qMobius-CFT-derivation}

In this section, we derive the exact CFT overlaps between undeformed states and their $q$-M\"obius-deformed counterparts, considering first the ground state and then general primary states.

\subsection{A. Deformed Hamiltonian and ground-state overlap}

We begin with the Virasoro generators satisfying the Virasoro algebra
\begin{align}
    [L_n,L_m] = (n-m)L_{n+m} + \frac{c}{12}n(n^2-1)\delta_{n+m,0},\quad n,m\in\mathbb{Z},
\end{align}
where the Virasoro generators are defined by the stress-energy tensor $T(x)$ and $\bar T(x)$
\begin{equation}\label{eq:App-Vir-Generator}
    L_n = \frac{L}{(2\pi)^2}\int_0^L \dd x  T(x)e^{i 2\pi n x/L}+\frac{c}{24}\delta_{n,0},\quad\quad\bar{L}_n = \frac{L}{(2\pi)^2}\int_0^L \dd x  \bar{T}(x)e^{-i 2\pi n x/L}+\frac{c}{24}\delta_{n,0}.
\end{equation}
Note that the CFT is defined on a compact system with periodic boundary condition $x\sim x+L$.
For a fixed positive integer $q$, we define
\begin{align}
    \widetilde{L}_0 = L_0+\frac{c}{24}(q^2-1).
\end{align}
Then, we have
\begin{align}
    [L_q-L_{-q},\widetilde{L}_0] = q(L_q+L_{-q}),
    \quad
    [L_q-L_{-q},L_q+L_{-q}] = 4q \widetilde{L}_0 ,
\end{align}
and the adjoint action of the unitary operator $e^{-\frac{\theta}{q}(L_q-L_{-q})}$ on $\widetilde{L}_0$ can be computed using the Baker-Campbell-Hausdorff formula:
\begin{align}
    e^{-\frac{\theta}{q}(L_q-L_{-q})}   \widetilde{L}_0  e^{\frac{\theta}{q}(L_q-L_{-q})}
    = \cosh(2\theta)\widetilde{L}_0 - \frac{\sinh(2\theta)}{2}(L_q+L_{-q}).
\label{eq:app-adjoint-action}
\end{align}

Using Eq.~\eqref{eq:app-adjoint-action}, we obtain the action of the unitary operator $U_q(\theta) = e^{-\frac{\theta}{q}(L_q-L_{-q})} e^{-\frac{\theta}{q}(\bar{L}_q - \bar{L}_{-q})}$ on the uniform Hamiltonian
\begin{align}
    H = \frac{2\pi}{L}
    \left( L_0 + \bar L_0 -\frac{c}{12} \right),
\label{eq:App-Uniform-H0}
\end{align}
which gives
\begin{align}
    U_q(\theta)   H   U^{\dagger}_q(\theta)
    = \frac{2\pi}{L} \left[\cosh(2\theta)\left( L_0 + \bar L_0 -\frac{c}{12} \right) - \frac{\sinh(2\theta)}{2} (L_q + L_{-q} + \bar{L}_q + \bar{L}_{-q}) \right]
    + \frac{\pi c q^2}{6L} [\cosh(2\theta) - 1] .
\end{align}
Rearranging the above expression immediately gives Eq.~(4) in the main text.

The deformed ground state is obtained by acting on the uniform CFT ground state $|0\rangle$ with the unitary transformation
\begin{align}
    \ket{0_q(\theta)} = U_q(\theta) |0\rangle = e^{-\frac{\theta}{q}(L_q-L_{-q})} e^{-\frac{\theta}{q}(\bar{L}_q - \bar{L}_{-q})} |0\rangle.
\label{eq:app-deformed-GS}
\end{align}
To obtain a normal-ordered form of the deformed ground state, we introduce
\begin{align}
    K_0=\frac{\widetilde{L}_0}{q},
    \quad
    K_+=\frac{L_{-q}}{q},
    \quad
    K_-=\frac{L_q}{q},
\end{align}
which satisfy the SU$(1,1)$ algebra
\begin{align}
    [K_0,K_\pm]=\pm K_\pm,
    \quad
    [K_-,K_+]=2K_0 .
\end{align}
Using the SU$(1,1)$ disentangling formula
\begin{align}
    e^{\eta(K_- - K_+)}
    = e^{-\tanh\eta K_+}
    e^{-2\ln(\cosh\eta)K_0}
    e^{\tanh\eta K_-},
\end{align}
with $\eta=-\theta$, we obtain
\begin{align}
    e^{-\frac{\theta}{q}(L_q-L_{-q})}\ket{0}
&=e^{\frac{\tanh \theta }{q}L_{-q}}e^{-\frac{2}{q}\ln (\cosh \theta )\left[
L_{0}+\frac{c}{24}(q^{2}-1)\right] }e^{-\frac{\tanh \theta }{q}
L_{q}}|0\rangle   \nonumber \\
&= e^{\frac{\tanh \theta }{q}L_{-q}}e^{-\frac{c}{12q}(q^{2}-1)\ln (\cosh
\theta )}|0\rangle   \nonumber \\
&= [\cosh \theta ]^{-\frac{c}{12q}(q^{2}-1)}e^{\frac{\tanh \theta }{q}
L_{-q}}|0\rangle,
\end{align}
where we have used $L_q\ket{0}=0$ for $q \geq 0$.
Including the antiholomorphic sector, the normalized $q$-M\"obius deformed ground state in Eq.~\eqref{eq:app-deformed-GS} is then given by
\begin{align}
    \ket{0_q(\theta)}
    =  [\cosh\theta]^{-\frac{c}{6q}(q^2-1)}
    e^{\frac{\tanh\theta}{q} (L_{-q}+\bar L_{-q})}\ket{0}.
\end{align}
The $q$-M\"obius ground state therefore takes the form
of an SU$(1,1)$ coherent state built on the conformal vacuum, and its overlap with the uniform ground state is
\begin{align}
    \braket{0}{0_q(\theta)}
    = [\cosh\theta]^{-\frac{c}{6q}(q^2-1)}.
\label{eq:app-CFT-overlap}
\end{align}
This yields Eqs.~(5) and (6) in the main text.

\subsection{B. Overlaps of deformed primary states}

We now generalize the construction to Virasoro primary states. A Virasoro primary state
$\ket{\phi}$ with conformal weight $(h,\bar{h})$ satisfies $L_0\ket{\phi} = h \ket{\phi}$, $\bar{L}_0\ket{\phi} = \bar{h} \ket{\phi}$, and $L_n\ket{\phi} = \bar{L}_n\ket{\phi} = 0$ for $n>0$. Using
\begin{align}
e^{-\frac{\theta }{q}(L_{q}-L_{-q})} \ket{\phi}
&= e^{\frac{\tanh \theta }{q}L_{-q}}e^{-\frac{2}{q}\ln (\cosh \theta )\left[
L_{0}+\frac{c}{24}(q^{2}-1)\right] }e^{-\frac{\tanh \theta }{q}L_{q}}\ket{\phi}   \nonumber \\
&= e^{\frac{\tanh \theta }{q}L_{-q}}e^{-\frac{2}{q}\ln (\cosh \theta )\left[
h+\frac{c}{24}(q^{2}-1)\right] } \ket{\phi}   \nonumber \\
&= [\cosh \theta ]^{-\frac{2}{q}[h+\frac{c}{24}(q^{2}-1)]}e^{\frac{\tanh
\theta }{q}L_{-q}}\ket{\phi},
\end{align}
we obtain the $q$-M\"obius deformed state
\begin{align}
    \ket{\phi_q(\theta)}  &= U_q(\theta) \ket{\phi}= [\cosh \theta ]^{-\frac{2}{q}[\Delta + \frac{c}{12}(q^{2}-1)]}     e^{\frac{\tanh\theta}{q} (L_{-q}+\bar L_{-q})}\ket{\phi},
\end{align}
where $\Delta = h + \bar{h}$ is the scaling dimension of $\ket{\phi}$. Then, we obtain the overlap
\begin{align}\label{eq:Primary-Overlap}
    \langle \phi | \phi_q(\theta)\rangle = [\cosh\theta]^{
-\frac{2\Delta}{q} -\frac{c(q^2-1)}{6q}}.
\end{align}
This shows that the overlap construction naturally extends from the conformal anomaly to the conformal data of primary states. When $q=1$, the central charge dependence in Eq.~\eqref{eq:Primary-Overlap} is removed, the state overlap only depends on the scaling dimension $\Delta$.

\section{II. Entanglement structure of the deformed ground state}
\label{app:qMobius-entanglement-spectrum}

In this section, we derive the reduced density operator and entanglement properties of the $q$-M\"obius ground state for a single interval $A=[x_1,x_2]$. Using the auxiliary cylinder coordinates, we first map the reduced-density-operator path integral to an annulus and determine its width $W_A^{(q,\theta)}(x_1,x_2)$, explaining how the endpoint Jacobian factors can produce non-monotonic, $q$-dependent spatial profiles. We then quantize the annulus in the open-string channel to obtain the reduced density operator, entanglement Hamiltonian, and entanglement spectrum, and show how normalization by the stress-tensor gap removes the geometric scale from the entanglement gaps. Finally, we derive the von Neumann entanglement entropy from the annulus partition function in the large-width limit, showing that the deformation enters through the annulus width while the universal coefficient remains fixed by the central charge.

\subsection{A. Conformal map and annulus width}

Let
\begin{equation}
    w=\tau+ix,
    \quad
    x\sim x+L,
    \quad
    \tau\in\mathbb{R},
\end{equation}
be the coordinate on the infinite Euclidean cylinder. 
The uniform CFT Hamiltonian on a circle of circumference $L$ is given by Eq.~\eqref{eq:App-Uniform-H0}.
At $\tau=0$, with $w=\tau+ix$ and $\bar w=\tau-ix$, one has $T(x)=T(w=ix)$ and $\bar T(x)=\bar T(\bar w=-ix)$ in Eq.~\eqref{eq:App-Vir-Generator}. Accordingly, the uniform ground state $\ket{0}$ is prepared by a path integral on this infinite cylinder.

For the $q$-M\"obius ground state, we introduce the auxiliary cylinder coordinates $\left(\xi(w),\bar{\xi}(\bar{w})\right)$ (a detailed derivation can be found in Appendix A of \cite{BaiC2024}):
\be
    \xi(w)=\frac{L_{\rm eff}}{2\pi}\ln\left(\frac{\cosh\theta -\sinh\theta e^{\frac{2q\pi w}{L}}}{\cosh\theta e^{\frac{2q\pi w}{L}}-\sinh\theta}\right)^{\frac{1}{q}},\quad\quad
    \bar{\xi}(\bar{w})=\frac{L_{\rm eff}}{2\pi}\ln\left(\frac{\cosh\theta -\sinh\theta e^{\frac{2q\pi \bar{w}}{L}}}{\cosh\theta e^{\frac{2q\pi \bar{w}}{L}}-\sinh\theta}\right)^{\frac{1}{q}},
\label{eq:app-auxiliary-cylinder}
\ee
where $L_{\rm eff}=L\cosh(2\theta)$.
This transformation maps the uniform infinite cylinder of circumference $L$ to the $q$-M\"obius infinite cylinder of circumference $L_{\rm eff}$. The $q$-M\"obius Hamiltonian generates time translations along this cylinder, i.e., 
\begin{align}
  H_q(\theta) &= \frac{2\pi}{L}
    \left[L_0 -\frac{\tanh(2\theta)}{2}
        \left(L_q+L_{-q}\right) \right.  \nonumber \left. +\bar L_0
        -\frac{\tanh(2\theta)}{2}
        \left(\bar L_q+\bar L_{-q}\right)
        -\frac{c}{12} \right].
\end{align}
Using the mode expansion in Eq.~\eqref{eq:App-Vir-Generator}, this Hamiltonian can be written as a spatially weighted integral of the stress tensor on the original cylinder, with deformation profile $f_{q,\theta}(x)=1-\tanh(2\theta)\cos\left(\frac{2q\pi x}{L}\right)$. The stress tensors in the coordinate $w$ and the $q$-M\"obius coordinate $\xi$ are related by
\begin{equation}
    T(w)=\left(\frac{\mathrm{d}\xi}{\mathrm{d}w}\right)^2T(\xi)+\frac{c}{12}\{\xi,w\},\quad\quad
    \bar{T}(\bar{w})=\left(\frac{\mathrm{d}\bar{\xi}}{\mathrm{d}\bar{w}}\right)^2\bar{T}(\bar{\xi})+\frac{c}{12}\{\bar{\xi},\bar{w}\}.
\end{equation}
On the equal-time slice $(w=ix, \bar w=-ix)$, the Jacobian is $(\mathrm{d}\xi/\mathrm{d}w)|_{w=ix}=-1/f_{q,\theta}(x)$. The Schwarzian derivative is
\begin{align}
    \{\xi,w\}|_{w=ix}=\frac{1}{2}\left[\left(\frac{2q\pi}{L_{\text{eff}}f_{q,\theta}(x)}\right)^2-\left(\frac{2q\pi}{L}\right)^2\right],
\end{align}
with the same expression for the antiholomorphic sector. Substituting these relations into the stress-tensor representation of $H_q(\theta)$, one obtains
\begin{equation}
    H_{q}(\theta)
    =
    \frac{2\pi}{L_{\rm eff}}
    \left(
        L_0^{(\xi)}+\bar{L}_0^{(\bar{\xi})}-\frac{c}{12}
    \right)+\frac{\pi cq^2}{6L_{\text{eff}}}-\frac{\pi cq^2}{6L}.
\label{eq:App-Uniform-H0-xi}
\end{equation}
Here, $(L_0^{(\xi)},\bar L_0^{(\bar{\xi})})$ are the zero modes defined on the $q$-M\"obius cylinder. Pulled back to the original coordinate $(w,\bar{w})$, they are
\begin{align}
    L_0^{(\xi)}&=\frac{L_{\text{eff}}}{(2\pi)^2}\int_0^L\dd x  f_{q,\theta}(x)\left[T(x)-\frac{c}{12}\{\xi(w),w\}|_{w=ix}\right]+\frac{c}{24},\\
    \bar{L}_0^{(\bar{\xi})}&=\frac{L_{\text{eff}}}{(2\pi)^2}\int_0^L\dd x  f_{q,\theta}(x)\left[\bar{T}(x)-\frac{c}{12}\{\bar{\xi}(\bar{w}),\bar{w}\}|_{\bar{w}=-ix}\right]+\frac{c}{24}.
\end{align}
Equation~\eqref{eq:App-Uniform-H0-xi} shows that the $q$-M\"obius Hamiltonian is equivalent to the uniform CFT Hamiltonian on a cylinder of circumference $L_{\rm eff}$, up to a constant Casimir-energy shift. Therefore, compared with the uniform ground state $\ket{0}$, the vacuum energy is shifted by
\begin{align}
    \Delta E_{\rm Casimir}=\frac{\pi cq^2}{6L_{\rm eff}}-\frac{\pi cq^2}{6L}.
\end{align}

Consider a subsystem
\begin{equation}
    A=[x_1,x_2]
\end{equation}
on the original equal-time slice $\tau=0$. It is easy to check that $\xi(ix_j)$ is a purely imaginary function; thus, we define the deformed spatial coordinate as
\be
    x^{\text{new}}(x)=-\text{Im}\left[\xi(ix)\right]=\frac{L_{\rm eff}}{q}\left[\frac{1}{\pi}\arctan\left(e^{2\theta}\tan\left[\pi\left(\frac{qx}{L}-\left\lfloor\frac{qx}{L} +\frac{1}{2}\right\rfloor\right)\right]\right)+\left\lfloor\frac{qx}{L} +\frac{1}{2}\right\rfloor\right],
\label{eq:app-deformed-coordinate}
\ee
where the floor function $\left\lfloor\frac{qx}{L} +\frac{1}{2}\right\rfloor$ corresponds to a specific choice of branch cut for $\xi$. This ensures that, when evaluated numerically in Python, $x^{\text{new}}(x)$ is a monotonically increasing function for $x\in[0,L]$, with $x^{\rm new}(0)=0$ and $x^{\rm new}(L)=L_{\text{eff}}$ (see Eq.~(10) in the main text). Accordingly, the endpoints are given by
\begin{equation}
    x_1^{\text{new}}=x^{\text{new}}(x_1),
    \quad
    x_2^{\text{new}}=x^{\text{new}}(x_2).
\end{equation}
The reduced density operator $\rho_A$ is represented by a path integral on the
auxiliary cylinder with a slit along the interval
$[x_1^{\text{new}},x_2^{\text{new}}]$ at $\tau=0$.

We first map the $q$-M\"obius cylinder to the complex plane by
\begin{equation}
    z=\exp\left(\frac{2\pi \xi}{L_{\rm eff}}\right).
\label{eq:app-cylinder-to-plane}
\end{equation}
The two endpoints become
\begin{equation}
    z_j
    =
    \exp\left(
        i\frac{2\pi  x_j^{\text{new}}}{L_{\rm eff}}
    \right),
    \quad
    j=1,2.
\end{equation}
Their separation on the unit circle is
\begin{equation}
    |z_1-z_2|
    =
    2\left|
    \sin\left(
        \frac{\pi(x_1^{\text{new}}-x_2^{\text{new}})}{L_{\rm eff}}
    \right)
    \right|.
\label{eq:app-z-separation}
\end{equation}

The slit plane is then uniformized to an annulus by
\begin{equation}
    \zeta
    =
    \ln\left(
        \frac{z-z_1}{z-z_2}
    \right),
    \quad
    \zeta=u+iv .
\label{eq:app-plane-to-annulus}
\end{equation}
The coordinate $v$ is periodic,
\begin{equation}
    v\sim v+2\pi,
\end{equation}
and the two regulated endpoints become the two annulus boundaries.  The annulus
width is the logarithmic distance between these two boundaries,
\begin{equation}
    W_A^{(q,\theta)}(x_1,x_2)
    =
    \ln\left(
        \frac{|z_1-z_2|^2}{\delta_1\delta_2}
    \right),
\label{eq:app-W-def}
\end{equation}
where $\delta_j$ is the ultraviolet (UV) cutoff around $z_j$ in the plane.

We now express $\delta_j$ in terms of the original cutoff $\epsilon$ in the
$x$ coordinate.  Since
\begin{equation}
    \mathrm{d} x^{\text{new}}
    =
    \frac{\mathrm{d}x}{f_{q,\theta}(x)},
\end{equation}
a cutoff $\epsilon$ in the original coordinate becomes
\begin{equation}
    \epsilon_j^{\text{new}}
    =
    \frac{\epsilon}{f_{q,\theta}(x_j)}
\end{equation}
in the auxiliary cylinder coordinate.  Under the map
$z=\exp(2\pi\xi/L_{\rm eff})$, the local cutoff becomes
\begin{equation}
    \delta_j
    =
    \left|
        \frac{\mathrm{d}z}{\mathrm{d}\xi}
    \right|_{\xi=i x_j^{\text{new}}}
    \epsilon_j^{\text{new}}
    =
    \frac{2\pi}{L_{\rm eff}}
    \frac{\epsilon}{f_{q,\theta}(x_j)}.
\label{eq:app-delta-j}
\end{equation}
Substituting Eqs.~\eqref{eq:app-z-separation} and
\eqref{eq:app-delta-j} into Eq.~\eqref{eq:app-W-def}, we obtain Eq.~(11) in the main text:
\begin{equation}
    W_A^{(q,\theta)}(x_1,x_2)
    =  \ln\left[
        \sin^2\left(
        \frac{\pi[x^{\text{new}}(x_1)-x^{\text{new}}(x_2)]}
             {L_{\rm eff}}
        \right)
    \right] + \ln\left[f_{q,\theta}(x_1)f_{q,\theta}(x_2)
        \frac{L_{\rm eff}^2}{\pi^2\epsilon^2}
        \right].
\label{eq:app-W-qMobius}
\end{equation}

By choosing $A=[0,x]$, the $x$-dependent part of the annulus width can be written as 
\begin{equation} 
W_A^{(q,\theta)}(x,0)=2\ln\left| \sin\left( \frac{\pi x^{\rm new}(x)}{L_{\rm eff}} \right) \right| +\ln f_{q,\theta}(x)+\mathrm{const.}, \end{equation} 
and therefore 
\begin{equation} 
\frac{\mathrm d W_A^{(q,\theta)}(x,0)}{\mathrm d x}=\frac{2\pi}{L_{\rm eff}f_{q,\theta}(x)} \cot\left( \frac{\pi x^{\rm new}(x)}{L_{\rm eff}} \right) + \frac{\partial_xf_{q,\theta}(x)}{f_{q,\theta}(x)}. 
\end{equation} 
Thus, although $\mathrm d x^{\rm new}/\mathrm d x=1/f_{q,\theta}(x)>0$, the annulus width need not be monotonic because of the endpoint-Jacobian contribution. The $q$-dependent modulation originates directly from the endpoint-Jacobian term $\ln f_{q,\theta}(x)$.
$f_{q,\theta}(x)$ contains $q$ oscillations over the full interval $[0,L]$, with spatial period $L/q$. For $\theta>0$, it is minimal at $x=mL/q$ and maximal at $x=(m+1/2)L/q$, where $m\in\mathbb{Z}$. As $\ln f_{q,\theta}(x)$ enters $W_A^{(q,\theta)}(x,0)$ additively, it locally decreases the annulus width near the minima of $f_{q,\theta}(x)$ and increases it near its maxima. This produces the $q$-dependent sequence of local dips and peaks in the entanglement entropy. The resulting profile is only periodic-like, rather than exactly periodic, because the $x^{\text{new}}$-dependent contribution, $2\ln\left|
\sin\left(
\frac{\pi x^{\rm new}(x)}{L_{\rm eff}}
\right)
\right|$, also varies with the total interval length measured from the fixed endpoint at $x=0$.
For $\theta=0$, one has $L_{\rm eff}=L$,
$f_{q,\theta}(x)=1$, and $x^{\text{new}}(x)=x$, so
\begin{equation}
    W_A(q,0)=\ln\left[
        \frac{L^2}{\pi^2\epsilon^2}
        \sin^2\left(
            \frac{\pi(x_1-x_2)}{L}
        \right)
    \right],
\end{equation}
which is the standard annulus width for a single interval on a circle.

\subsection{B. Reduced density operator and entanglement Hamiltonian}

The annulus has coordinates
\begin{equation}
    \zeta=u+iv,
    \quad
    0\leq u\leq W_A^{(q,\theta)}(x_1,x_2),
    \quad
    v\sim v+2\pi .
\end{equation}
In the open-string channel \cite{DiFrancesco1997}, $v$ is Euclidean time and $u$ is the spatial coordinate on a strip of width $W_A^{(q,\theta)}(x_1,x_2)$. The open-string channel Hamiltonian is
\begin{equation}
    H_{\rm strip}
    =
    \frac{\pi}{W_A^{(q,\theta)}(x_1,x_2)}
    \left(
        L_0^{\rm op.}-\frac{c}{24}
    \right).
\label{eq:app-H-strip}
\end{equation}
Here $L_0^{\rm op.}$ is the Virasoro zero mode in the open-string channel. The Casimir shift is $-c/24$, rather than $-c/12$, because the open-string channel Hilbert space carries a single Virasoro algebra.

Going once around the annulus corresponds to Euclidean time evolution by
\begin{equation}
    \Delta v=2\pi .
\end{equation}
Therefore the unnormalized reduced density operator is
\begin{equation}
\begin{split}
    \rho_A^{\rm unnorm}
    &=
    e^{-2\pi H_{\rm strip}}\\
    &=
    \exp\left[
        -\frac{2\pi^2}{W_A^{(q,\theta)}(x_1,x_2)}
        \left(
            L_0^{\rm op.}-\frac{c}{24}
        \right)
    \right].
\end{split}   
\end{equation}
After imposing $\tr\rho_A=1$, we obtain
\begin{equation}
    \rho_A(q,\theta)
    =
    \frac{1}{Z_A}
    \exp\left[
        -\frac{2\pi^2}{W_A^{(q,\theta)}(x_1,x_2)}
        \left(
            L_0^{\rm op.}-\frac{c}{24}
        \right)
    \right],
\label{eq:app-rhoA}
\end{equation}
where
\begin{equation}
    Z_A
    =
    \tr
    \exp\left[
        -\frac{2\pi^2}{W_A^{(q,\theta)}(x_1,x_2)}
        \left(
            L_0^{\rm op.}-\frac{c}{24}
        \right)
    \right].
\end{equation}
Equivalently,
\begin{equation}
    \rho_A(q,\theta)
    =
    \frac{1}{Z_A}
    q_A^{ L_0^{\rm op.}-c/24},
    \quad
    q_A
    =
    \exp\left[
        -\frac{2\pi^2}{W_A^{(q,\theta)}(x_1,x_2)}
    \right].
\end{equation}

\subsection{C. Entanglement spectrum and normalized gaps}

Let the annulus Hilbert space be decomposed into eigenstates of
$L_0^{\rm op.}$,
\begin{equation}
    L_0^{\rm op.}|\alpha\rangle
    =
    h_\alpha|\alpha\rangle,
\end{equation}
where $\ket{\alpha}$ is a state with conformal weight $h_{\alpha}$.
Then, Eq.~\eqref{eq:app-rhoA} gives the eigenvalues of the reduced density
matrix,
\begin{equation}
    \lambda_\alpha
    =
    \frac{1}{Z_A}
    \exp\left[
        -\frac{2\pi^2}{W_A^{(q,\theta)}(x_1,x_2)}
        \left(
            h_\alpha-\frac{c}{24}
        \right)
    \right].
\label{eq:app-lambda-alpha}
\end{equation}

The entanglement energies are defined by
\begin{equation}
    \varepsilon_\alpha
    =
    -\ln\lambda_\alpha .
\end{equation}
Therefore
\begin{equation}
    \varepsilon_\alpha
    =
    \ln Z_A
    +
    \frac{2\pi^2}{W_A^{(q,\theta)}(x_1,x_2)}
    \left(
        h_\alpha-\frac{c}{24}
    \right).
\label{eq:app-epsilon-alpha}
\end{equation}
Taking gaps relative to the lowest entanglement level removes both the
normalization $\ln Z_A$ and the Casimir shift:
\begin{equation}
    \varepsilon_\alpha-\varepsilon_0
    =
    \frac{2\pi^2}{W_A^{(q,\theta)}(x_1,x_2)}
    \left(
        h_\alpha-h_0
    \right),
\label{eq:app-entanglement-gap}
\end{equation}
which yields Eq.~(12) in the main text.
Thus, the $q$-M\"obius deformation changes the overall entanglement scale through the annulus width $W_A^{(q,\theta)}(x_1,x_2)$, but the low-lying entanglement gaps are still organized by conformal weights in the annulus channel.

To remove the geometric scale, we normalize entanglement gaps by the stress-tensor gap. The (holomorphic) stress tensor $T$ has conformal weight $h_T=2$. Therefore,
\begin{equation}
    \varepsilon_T-\varepsilon_0
    =
    \frac{2\pi^2}{W_A^{(q,\theta)}(x_1,x_2)}
    \left(
        h_T-h_0
    \right).
\end{equation}
In the vacuum-dominant channel, $h_0=0$, such that
\begin{equation}
    \varepsilon_T-\varepsilon_0
    =
    \frac{4\pi^2}{W_A^{(q,\theta)}(x_1,x_2)} .
\end{equation}
Combining this with Eq.~\eqref{eq:app-entanglement-gap}, we obtain
\begin{equation}
    \frac{
        \varepsilon_\alpha-\varepsilon_0
    }{
        \varepsilon_T-\varepsilon_0
    }
    =
    \frac{h_\alpha}{2}.
\label{eq:app-normalized-ES}
\end{equation}
More generally, if the lowest state in the relevant annulus channel has
$h_0\neq0$, then
\begin{equation}
    \frac{
        \varepsilon_\alpha-\varepsilon_0
    }{
        \varepsilon_T-\varepsilon_0
    }
    =
    \frac{
        h_\alpha-h_0
    }{
        h_T-h_0
    }.
\label{eq:app-normalized-ES-general}
\end{equation}

\subsection{D. Entanglement entropy}

In the end, we derive the von Neumann entanglement entropy from the entanglement energies. From Eq.~\eqref{eq:app-rhoA}, it is useful to define
\begin{equation}
    \beta_A
    =
    \frac{2\pi^2}{W_A^{(q,\theta)}(x_1,x_2)} .
\end{equation}
Then the reduced density operator takes the thermal form
\begin{equation}
    \rho_A(q,\theta)
    =
    \frac{1}{Z_A(\beta_A)}
    \exp\left[
        -\beta_A
        \left(
            L_0^{\rm op.}-\frac{c}{24}
        \right)
    \right],
\end{equation}
with
\begin{equation}
    Z_A(\beta_A)
    =Z_A=
    \tr_{\mathcal{H}_{\rm op}}
    \exp\left[
        -\beta_A
        \left(
            L_0^{\rm op.}-\frac{c}{24}
        \right)
    \right].
\end{equation}
Here the trace is over the full open-string channel Hilbert space, including both primary states and their descendants. The von Neumann entropy is therefore
\begin{equation}
\begin{split}
    S_A
    &=
    -\tr (\rho_A \ln \rho_A)
    =
    \sum_\alpha \lambda_\alpha \varepsilon_\alpha  \\
    &=
    \ln Z_A(\beta_A)
    +
    \beta_A
    \left\langle
        L_0^{\rm op.}-\frac{c}{24}
    \right\rangle_{\rho_A}.
\end{split}
\end{equation}
Using
\begin{equation}
    \partial_{\beta_A}\ln Z_A(\beta_A)
    =
    -
    \left\langle
        L_0^{\rm op.}-\frac{c}{24}
    \right\rangle_{\rho_A},
\end{equation}
we obtain the standard thermodynamic identity
\begin{equation}
    S_A
    =
    \left(
        1-\beta_A\partial_{\beta_A}
    \right)
    \ln Z_A(\beta_A).
\label{eq:app-EE-thermal-identity}
\end{equation}

To evaluate $Z_A(\beta_A)$ in the universal regime, we take the UV limit $\epsilon\to0$, for which
\begin{equation}
   W_A^{(q,\theta)}(x_1,x_2)\gg1,
    \quad
    \beta_A=\frac{2\pi^2}{W_A^{(q,\theta)}(x_1,x_2)}\ll1 .
\end{equation}
The open-string channel annulus partition function is then evaluated by a modular transformation to the closed channel. The leading contribution is dominated by the ground state, giving
\begin{equation}
    \ln Z_A(\beta_A)
    =
    \frac{\pi^2 c}{6\beta_A}
    +
    \ln(g_a g_b)
    +
    s_{\rm UV},
\end{equation}
where $g_a$ and $g_b$ are Affleck-Ludwig boundary $g$ factors~\cite{Affleck1991} associated with the entangling cuts, and $s_{\rm UV}$ is a nonuniversal UV-dependent constant. Applying
Eq.~\eqref{eq:app-EE-thermal-identity}, we find
\begin{equation}
    S_A
    =
    \left(
        1-\beta_A\partial_{\beta_A}
    \right)
    \left[
        \frac{\pi^2 c}{6\beta_A}
        +
        \ln(g_a g_b)
        +
        s_{\rm UV}
    \right]  
    =
    \frac{\pi^2 c}{3\beta_A}
    +
    s_1,
\end{equation}
where
\begin{equation}
    s_1
    =
    \ln(g_a g_b) + s_{\rm UV}
\end{equation}
is a nonuniversal additive constant. Since
$\beta_A=2\pi^2/W_A^{(q,\theta)}(x_1,x_2)$, this becomes
\begin{equation}
    S_A
    =
    \frac{c}{6}W_A^{(q,\theta)}(x_1,x_2)
    +
    s_1.
\label{eq:app-EE-WA}
\end{equation}
Substituting Eq.~\eqref{eq:app-W-qMobius}, we finally obtain Eq.~(13) in the main text
\begin{equation}
    S_A =     \frac{c}{3}
    \ln\left|
        \sin\left(
        \frac{
            \pi[x^{\text{new}}(x_1)-x^{\text{new}}(x_2)]
        }{
            L_{\rm eff}
        }
        \right) \right|
    +
    \frac{c}{6}
    \ln\left[
        f_{q,\theta}(x_1)f_{q,\theta}(x_2)
        \frac{L_{\rm eff}^2}{\pi^2\epsilon^2}
    \right]
    +  
    s_1 .
\label{eq:app-qMobius-EE}
\end{equation}
Thus, the $q$-M\"obius deformation modifies the entanglement entropy through the deformed annulus width $W_A^{(q,\theta)}(x_1,x_2)$, while the universal coefficient of the logarithm remains fixed by the central charge.

\section{III. Lattice models and numerical benchmarks}
\label{app:numerical}

In this section, we provide additional numerical results supporting the main text. Before presenting these results, we summarize the microscopic lattice models used in the numerical calculations. Throughout this section, periodic boundary conditions are assumed. The corresponding $q$-M\"obius deformed Hamiltonian is obtained by multiplying each local Hamiltonian term by the envelope function $f_{q,\theta}(X)$ defined in the main text.

For the transverse-field Ising chain, whose critical point is described by the Ising CFT with central charge $c=1/2$, we use
\begin{equation}
    H_{\mathrm{Ising}}=-\sum_{j=1}^{L}\left(X_jX_{j+1}+hZ_j\right),\quad h=1,
\label{eq:app-Ising-Hamiltonian}
\end{equation}
where $X_j$ and $Z_j$ are Pauli operators.

For the antiferromagnetic spin-$1/2$ Heisenberg chain, whose low-energy theory is the $\mathrm{SU}(2)_1$ Wess-Zumino-Witten CFT with central charge $c=1$, we use
\begin{equation}
H_{\mathrm{Heis}}=\sum_{j=1}^{L}\left(X_jX_{j+1}+Y_jY_{j+1}+Z_jZ_{j+1}\right),
\label{eq:app-Heisenberg-Hamiltonian}
\end{equation}
where $X_j,Y_j,Z_j$ are Pauli operators. 

We also consider the Blume-Capel model~\cite{Blume1966,Capel1966,Xavier2012},
\begin{equation}
    H_{\mathrm{BC}}=-\sum_{j=1}^{L}\left[S_j^zS_{j+1}^z-\gamma S_j^x-\delta (S_j^z)^2\right],
\label{eq:app-BC-Hamiltonian}
\end{equation}
where $S_j^x$ and $S_j^z$ are spin-1 operators. 
At $\gamma=0.41563$ and $\delta=0.91024$, the model is tuned to the tricritical Ising fixed point with central charge $c=7/10$.

The quantum three-state Potts chain is defined as~\cite{Albertini1992,Gomez1996}
\begin{equation}
    H_{\mathrm{Potts}}=-\sum_{j=1}^{L}\left(JX_jX_{j+1}^{\dagger}+gZ_j+\mathrm{h.c.}\right),
\label{eq:app-Potts-Hamiltonian}
\end{equation}
where $X_j^3=Z_j^3=1$, $X_j^2=X_j^\dagger$, $Z_j^2=Z_j^\dagger$, and $Z_jX_j=\omega X_jZ_j$ with $\omega=e^{2\pi {\rm i}/3}$. 
Explicitly, we take $Z=\mathrm{diag}(1,\omega,\omega^2)$ and $X_{ab}=\delta_{a+1,b}$, with indices understood modulo three. 
For $J=g=1$, the model is at the ferromagnetic critical point described by the $\mathbb{Z}_3$ parafermion CFT with central charge $c=4/5$~\cite{Dotsenko1984}. 
We also consider the antiferromagnetic self-dual point $J=g=-1$, whose continuum limit is described by the $\text{U}(1)_6$ CFT with central charge $c=1$~\cite{Kedem1993,LiW2015}.

\begin{figure}
    \centering
    \includegraphics[width=0.85\linewidth]{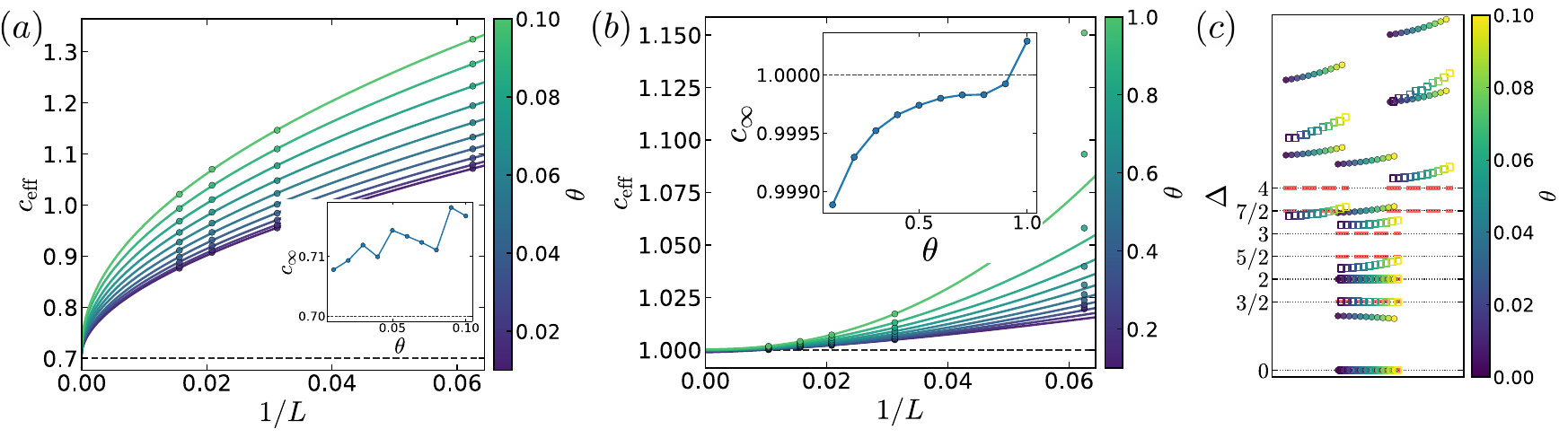}
    \caption{Central-charge extraction and entanglement-spectrum structure in additional deformed critical chains.
    (a,b) Finite-size extrapolation of the effective central charge $c_{\mathrm{eff}}$ for the deformed (a) Blume-Capel and (b) antiferromagnetic three-state Potts models.
    The main panels show $c_{\mathrm{eff}}$ as a function of $1/L$ for different deformation strengths $\theta$, with colored markers denoting numerical data and solid curves denoting finite-size fits of the form $aL^{-p}+b$ based on the four largest system sizes.
    The insets display the thermodynamic-limit estimates $c_{\infty}$ for different $\theta$ obtained from the $L\rightarrow\infty$ extrapolation.
    Dashed horizontal lines indicate the expected CFT central charges, $c=7/10$ for the Blume-Capel model and $c=1$ for the antiferromagnetic Potts model.
    DMRG calculations are performed with bond dimension $\chi=1200$ for $L=16,32,48,64$ in (a) and $\chi=800$ for $L=16,32,48,64,96$ in (b).
    (c) Rescaled entanglement spectrum of the deformed Blume-Capel model.
    The vertical axis shows the scaling dimensions $\Delta$ extracted from the entanglement spectrum, normalized by fixing the stress-tensor level to $\Delta=2$.
    Colors indicate the deformation strength $\theta$.
    Filled circles show numerical results for $L=64$, while open squares represent thermodynamic-limit estimates obtained from finite-size fits of the form $aL^{-p}+b$ based on the four largest system sizes.
    Red dashed lines mark the conformal tower levels predicted by the tricritical Ising CFT with free boundary conditions.}
    \label{fig:ceff of BC AFMPotts}
\end{figure}

\begin{figure}
    \centering
    \includegraphics[width=0.5\linewidth]{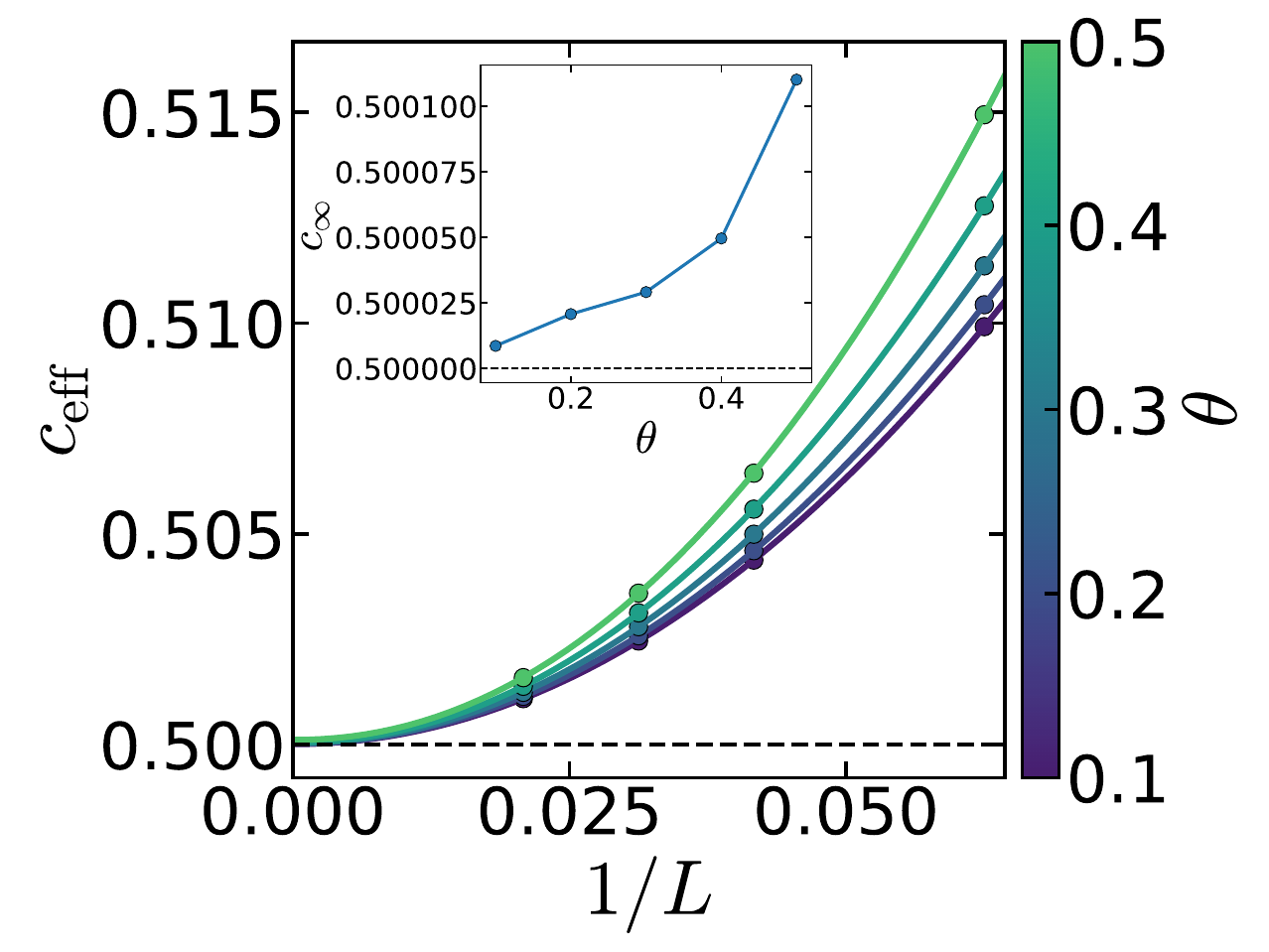}
    \caption{Central charge extraction for the critical cluster-Ising model with three-site interactions. The effective central charge $c_{\rm eff}$ is plotted against $1/L$ for system sizes up to $L=48$ and deformation parameters $\theta \in \{0.1, 0.2, 0.3, 0.4, 0.5\}$. Curves denote finite-size extrapolations, and the horizontal dashed line marks the exact Ising CFT value ($c=1/2$). The extrapolated results are highly consistent with the analytical prediction.
    }
    \label{fig:clusterIsing}
\end{figure}

Next, we consider the SU(3) Uimin-Lai-Sutherland chain~\cite{Uimin1970,LaiCK1974,Sutherland1975},
\begin{equation}
 H_{\mathrm{SU(3)}}=J\sum_{j=1}^{L}\sum_{a=1}^{8}T_j^aT_{j+1}^a,
\label{eq:app-SU3-Hamiltonian}
\end{equation}
where $T^a=\lambda^a/2$ and $\lambda^a$ are the eight Gell-Mann matrices. 
This model is equivalent, up to an additive constant and an overall normalization, to a spin-1 bilinear-biquadratic chain
\begin{equation}
H_{\mathrm{ULS}}=J_{\mathrm{ULS}}\sum_{j=1}^{L}\left[\mathbf{S}_j\cdot\mathbf{S}_{j+1}+\left(\mathbf{S}_j\cdot\mathbf{S}_{j+1}\right)^2\right].
\label{eq:app-ULS-Hamiltonian}
\end{equation}
The equivalence follows from the identity
\begin{equation}
    P_{j,j+1}=\frac{1}{3}+\frac{1}{2}\sum_{a=1}^{8}\lambda_j^a\lambda_{j+1}^a=\mathbf{S}_j\cdot\mathbf{S}_{j+1}+\left(\mathbf{S}_j\cdot\mathbf{S}_{j+1}\right)^2-1,
\label{eq:app-SU3-permutation}
\end{equation}
where $P_{j,j+1}$ is the permutation operator between neighboring SU(3) spins in the fundamental representation. 
The low-energy theory of this chain is the $\mathrm{SU}(3)_1$ Wess-Zumino-Witten CFT with central charge $c=2$~\cite{Affleck1988,Fuehringer2008,Aguado2009}.

Finally, we explore the method's applicability for the critical cluster-Ising model, which features a three-site interaction. The Hamiltonian is
\begin{equation}
    H_{\text{CI}}=-\sum_{j=1}^L(X_jX_{j+1}+X_jZ_{j+1}X_{j+2}),   
\end{equation}
where $X_j$ and $Z_j$ are Pauli operators. Its low-energy theory is the $c=1/2$ Ising CFT. This example supports that our prescription applies to models with multi-site interactions.

\begin{figure}
    \centering
    \includegraphics[width=1.0\linewidth]{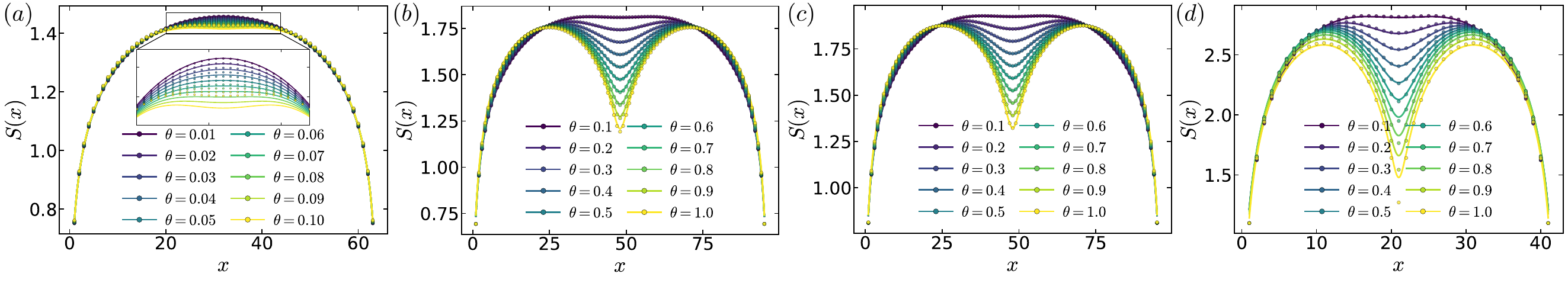}
    \caption{Entanglement-entropy profiles of deformed critical chains.
    Panels show (a) the Blume-Capel, (b) spin-$\frac{1}{2}$ Heisenberg, (c) antiferromagnetic three-state Potts, and (d) SU(3) Uimin-Lai-Sutherland chains.
    The entanglement entropy $S(x)$ is plotted as a function of the bipartition position $x$ for different deformation strengths $\theta$, with $L=64$ in (a), $L=96$ in (b,c), and $L=42$ in (d).
    Colored markers show numerical data, and solid curves show the analytical predictions from Eq.~\eqref{eq:app-qMobius-EE}.
    In panels (a)-(c), a single nonuniversal UV constant is used for all $\theta$, whereas in panel (d) the UV constant is allowed to vary with $\theta$.
    The inset in (a) enlarges the central region to show the weak-deformation behavior.}
    \label{fig:EE of BC XXX AFMPotts SU3}
\end{figure}

\subsection{A. Extraction of central charge and entanglement entropy}

After defining these microscopic models, we first present additional numerical checks of the central charge and entanglement results discussed in the main text. 
Figure~\ref{fig:ceff of BC AFMPotts}(a,b) shows the finite-size extrapolation of the effective central charge for the deformed Blume-Capel model at its tricritical point and for the antiferromagnetic three-state Potts model. In addition, Fig.~\ref{fig:clusterIsing} shows the finite-size extrapolation of the effective central charge for the cluster-Ising chain.
The extrapolated values converge to the expected CFT central charges $c=7/10$, $c=1$, and $c=1/2$, respectively, demonstrating that the ground-state-overlap estimator applies beyond the representative models presented in the main text.
Figure~\ref{fig:ceff of BC AFMPotts}(c) shows the rescaled entanglement spectrum of the deformed Blume-Capel model. 
After normalizing the spectrum by fixing the stress-tensor level to $\Delta=2$, the low-lying levels agree with the conformal tower structure of the tricritical Ising CFT with free boundary conditions. 
This provides an additional check that the $q$-M\"obius deformation preserves the universal boundary-CFT structure of the entanglement spectrum.

In Fig.~\ref{fig:EE of BC XXX AFMPotts SU3}, we further compare the entanglement-entropy profiles of several deformed critical chains with the analytical prediction in Eq.~\eqref{eq:app-qMobius-EE}. 
The agreement confirms that the deformation modifies the entanglement primarily through the conformal reparametrization and the associated endpoint Jacobian factors. 

\subsection{B. Extraction of primary field scaling dimensions}

We now apply Eq.~\eqref{eq:Primary-Overlap} to three representative examples to demonstrate the extraction of primary-field scaling dimensions. 
Numerically, combining Eqs.~\eqref{eq:app-CFT-overlap} and \eqref{eq:Primary-Overlap} gives
\begin{equation}
\label{eq:scaling_dimension_numerical}
    \Delta_n=-\frac{q}{2\ln(\cosh\theta)}\ln\left|\frac{\langle \phi_n|\phi_n(q,\theta)\rangle}{\langle 0|0_q(\theta)\rangle}\right|.
\end{equation}
Thus, the scaling dimension of a primary field can be directly extracted from the ratio between the corresponding excited-state overlap and the ground-state overlap.

To illustrate the extraction of scaling dimensions via Eq.~\eqref{eq:scaling_dimension_numerical}, we analyze three representative critical models, beginning with the Ising Hamiltonian in Eq.~\eqref{eq:app-Ising-Hamiltonian}. We focus on the two lowest excited states corresponding to the primary fields $\sigma$ and $\epsilon$, whose scaling dimensions are $\Delta_\sigma=1/8$ and $\Delta_\epsilon=1$. By employing DMRG, we obtain the respective excited states for both the original and deformed Hamiltonians and calculate their overlaps. As shown in Fig.~\ref{fig:IsingPrimary}, the numerically extracted scaling dimensions, after finite-size extrapolation, are in remarkable agreement with the exact CFT values. We emphasize that this high level of accuracy is achieved even for moderately sized systems of $L=48$.
    
    \begin{figure}
        \centering
        \includegraphics[width=0.7\linewidth]{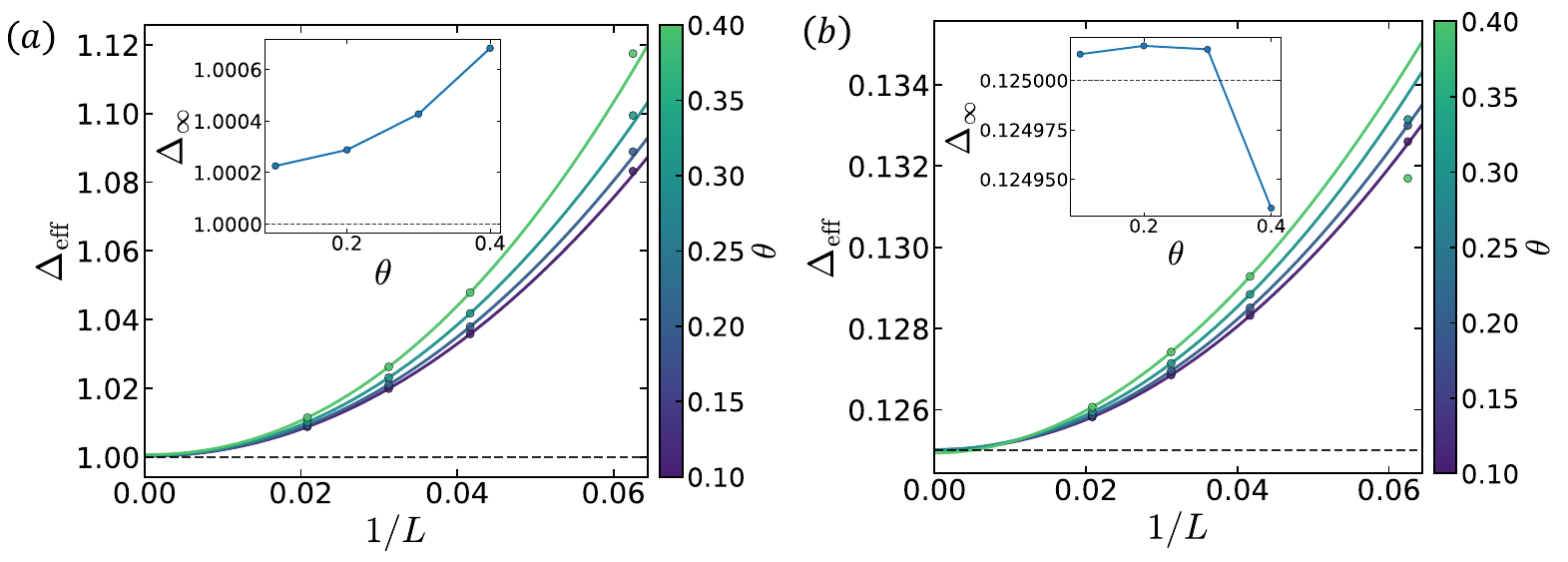}
        \caption{
        Numerical extraction of primary-field scaling dimensions in the critical transverse-field Ising model. (a), (b) Finite-size estimates of $\Delta_\epsilon$ and $\Delta_\sigma$ plotted against $1/L$ for system sizes $L \in \{16, 24, 32, 48\}$ and deformation strengths $\theta \in \{0.1, 0.2, 0.3, 0.4\}$. Solid curves denote the finite-size extrapolations; horizontal dashed lines mark the exact CFT predictions ($\Delta_\epsilon=1$, $\Delta_\sigma=1/8$). The insets show the thermodynamic-limit extrapolated values $\Delta_\infty$ versus $\theta$, confirming excellent agreement with exact analytical results.
        }
        \label{fig:IsingPrimary}
    \end{figure}

    \begin{figure}
        \centering
        \includegraphics[width=0.7\linewidth]{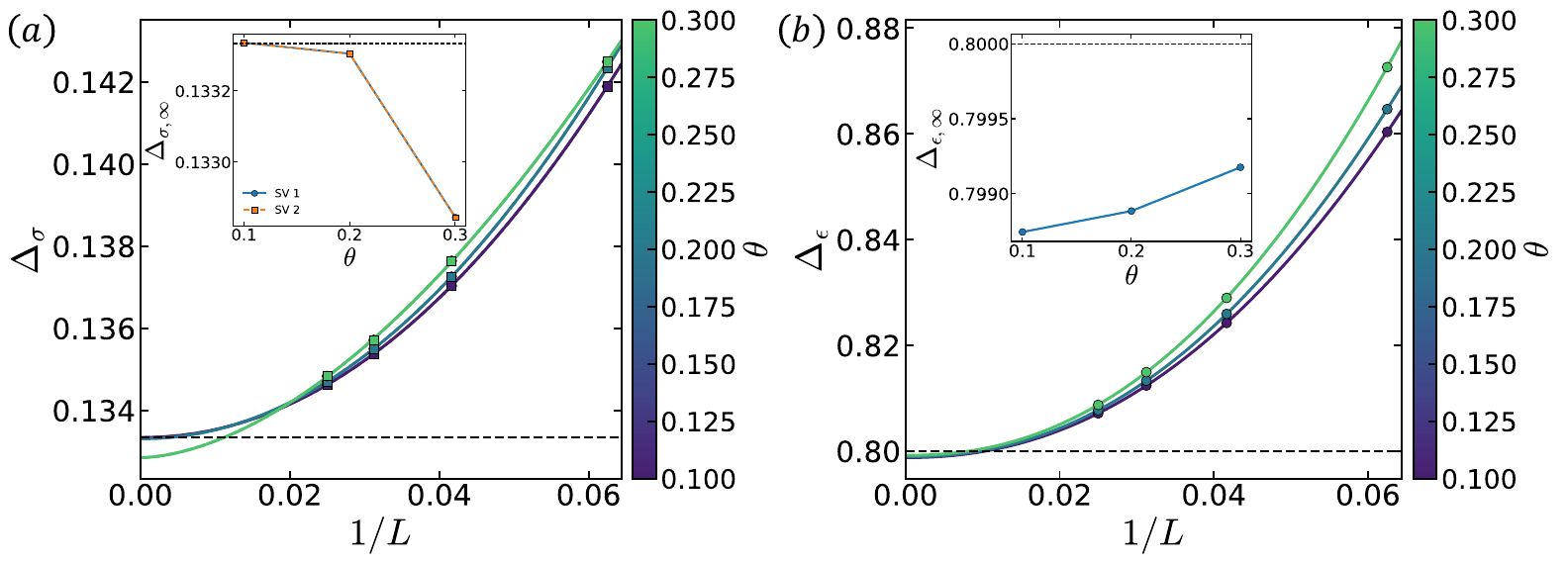}
        \caption{Numerical extraction of scaling dimensions in the critical three-state Potts model. Results are presented for system sizes $L \in \{16,24,32,40\}$ and deformation strengths $\theta \in \{0.1,0.2,0.3\}$. (a) Scaling dimensions in the $\sigma$ sector. Derived from the two eigenvalues of the $2\times 2$ overlap matrix between the degenerate excited-state manifolds of the original and deformed Hamiltonians, the nearly degenerate estimates are distinguished by square and circle markers. (b) The scaling dimension $\Delta_\epsilon$ obtained from the nondegenerate third excited state. In both panels, finite-size extrapolations converge cleanly to the exact CFT values ($\Delta_\sigma=2/15$, $\Delta_\epsilon=4/5$).
        }
        \label{fig:Pottsprimary}
    \end{figure}

As a second example, we consider the critical three-state Potts model, described by Eq.~\eqref{eq:app-Potts-Hamiltonian} with $J=g=1$. We focus on the three lowest-lying excited states. The first two states are degenerate and correspond to the primary field $\sigma$, with an exact scaling dimension of $\Delta_\sigma=2/15$, while the third corresponds to the primary field $\epsilon$, with a scaling dimension of $\Delta_\epsilon=4/5$. Because the individual eigenstates obtained from the original and deformed Hamiltonians are not uniquely defined within the two-dimensional degenerate subspace, we construct the $2\times 2$ overlap matrix between the corresponding excited-state manifolds. Diagonalizing this matrix yields two nearly degenerate eigenvalues, which provide two estimates for the scaling dimension in the $\sigma$ sector. In Fig.~\ref{fig:Pottsprimary}, these two estimates are plotted separately using square and circle markers. For the third excited state, which is nondegenerate, we extract $\Delta_\epsilon$ directly from the scalar overlap. The extracted scaling dimensions for system sizes $L=16, 24, 32, 40$ and deformation parameters $\theta=0.1, 0.2, 0.3$ are presented in Fig.~\ref{fig:Pottsprimary}. After finite-size extrapolation, the numerical results agree well with the exact CFT predictions for both primary fields.

As a final example, we analyze the spin-$1/2$ $J_1$-$J_2$ Heisenberg chain, $H=\sum_i\mathbf{S}_i\cdot\mathbf{S}_{i+1}+0.241\sum_i\mathbf{S}_i\cdot\mathbf{S}_{i+2}$. By tuning the next-nearest-neighbor coupling to $J_2 = 0.241$, we suppress the leading marginal perturbation and substantially reduce logarithmic finite-size corrections. Following the procedure outlined previously, we first evaluate the central charge via the ground-state overlap. We then target the lowest-energy excitation in the $S^z=1$ sector—corresponding to a primary field with scaling dimension $\Delta=1/2$—and extract its scaling dimension from the ratio of the excited-state to ground-state overlaps. As shown in Fig.~\ref{fig:HeisenbergPrimary}, the finite-size extrapolations for both quantities converge to the exact CFT predictions ($c=1$, $\Delta=1/2$). This demonstrates the robustness of our overlap approach in extracting primary-field scaling dimensions alongside the central charge.

    \begin{figure}
        \centering
        \includegraphics[width=0.7\linewidth]{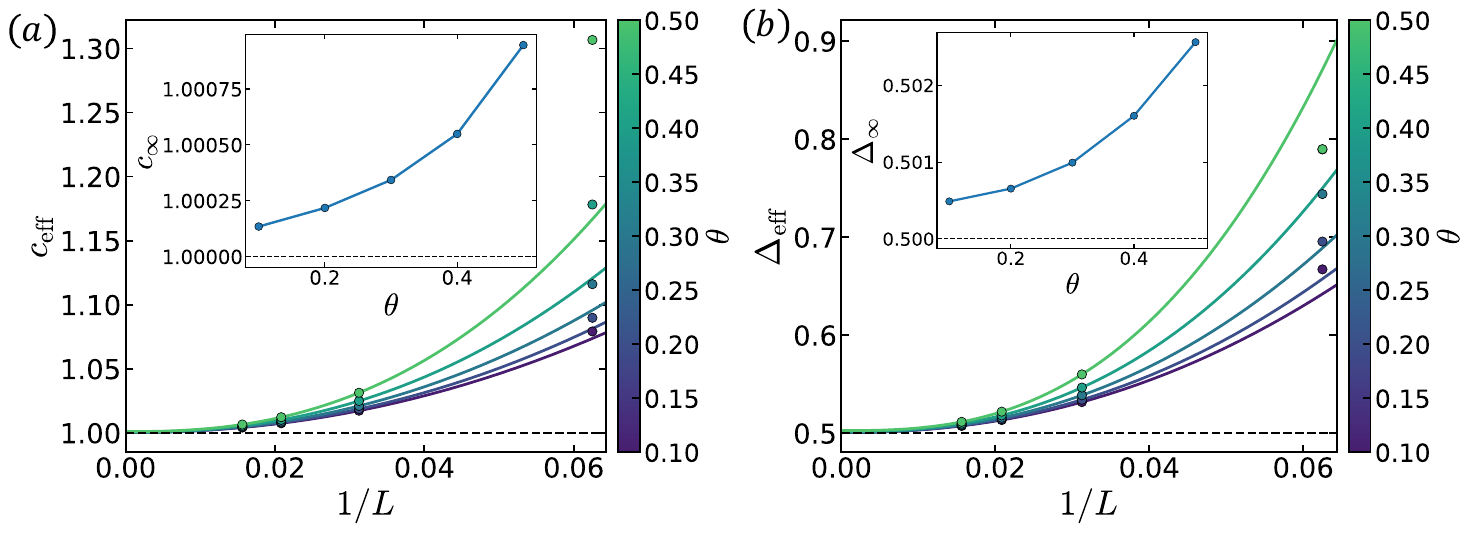}
        \caption{
        Conformal data extraction for the spin-$1/2$ $J_1$-$J_2$ Heisenberg chain at $J_2/J_1=0.241$. Finite-size estimates of (a) the effective central charge $c_{\rm eff}$ and (b) the effective scaling dimension $\Delta_{\rm eff}$ plotted against $1/L$. Results are shown for system sizes $L \in \{16, 32, 48, 64\}$ and deformation strengths $\theta \in \{0.1, 0.2, 0.3, 0.4, 0.5\}$. The scaling dimension is obtained from the lowest-energy excitation in the $S^z=1$ sector. Horizontal dashed lines denote the expected CFT values ($c=1$, $\Delta=1/2$). The insets show the thermodynamic-limit extrapolations $c_\infty$ and $\Delta_\infty$ versus $\theta$.
        }
        \label{fig:HeisenbergPrimary}
    \end{figure}

\section{IV. $q$-M\"obius deformation of CFT correlator wave functions and long-range parent Hamiltonians}
\label{app:HS-construction}

The equal-time geometry derived in Sec.~\ref{app:qMobius-entanglement-spectrum} admits an exact realization at the level of CFT-correlator wave functions. For a periodic chain of $L$ sites, let $s_j$ label a local basis state at site $j=1,\ldots,L$, and let $\varphi_{s_j}$ be the corresponding chiral primary field with chiral conformal weight $h_{s_j}$. Throughout this subsection, $\mathrm{i}$ denotes the imaginary unit, while italic $i,j,k,m,n$ label lattice sites. We use angle brackets to denote a fixed chiral conformal block, with the fields ordered by increasing site index. The corresponding uniform wave function, denoted by $\ket{\Psi_0}$, can be written as
\begin{equation}
    \ket{\Psi_0}
    \propto
    \sum_{\{s_j\}}
    \left\langle
    \prod_{j=1}^{L}
    \varphi_{s_j}\left(e^{\mathrm{i}\frac{2\pi j}{L}}\right)
    \right\rangle
    \ket{s_1,\ldots,s_L}.
\label{eq:app-uniform-IDMPS}
\end{equation}
The proportionality constant normalizes the state. Equation~\eqref{eq:app-uniform-IDMPS} is an infinite-dimensional matrix product state (IDMPS): the primary fields serve as site-dependent matrices acting successively on the generally infinite-dimensional chiral CFT Hilbert space~\cite{CiracSierra2010}. Fixed local basis phases may multiply the coefficients in Eq.~\eqref{eq:app-uniform-IDMPS}. Since these phases do not depend on the insertion coordinates and remain unchanged under the deformation, we leave them implicit at this stage.

The Virasoro unitary $U_q(\theta)$ in Eq.~\eqref{eq:app-deformed-GS}, where the positive integer $q$ fixes the modulation wave number and the real parameter $\theta$ fixes its strength, generates the finite conformal transformation in Eq.~\eqref{eq:app-auxiliary-cylinder}. Here $w$ is the original cylinder coordinate, $\xi(w)$ is the transformed cylinder coordinate, and $L_{\rm eff}$ is the corresponding effective circumference. Exponentiating that transformation gives
\begin{equation*}
    \exp\left[-\frac{2\pi q}{L_{\rm eff}}\xi(w)\right]
    =
    \frac{\cosh\theta e^{\frac{2\pi q w}{L}}-\sinh\theta}
    {\cosh\theta-\sinh\theta e^{\frac{2\pi q w}{L}}}.
\end{equation*}
On the equal-time slice, $\xi(\mathrm{i}j)=-\mathrm{i}x^{\text{new}}(j)$, where $x^{\text{new}}(j)$ is the deformed spatial coordinate defined in Eq.~\eqref{eq:app-deformed-coordinate}. Substituting $w=\mathrm{i}j$ sends the uniform insertion at site $j$ to the plane coordinate $z_j$,
\begin{equation}
    z_j
    =
    \exp\left[
    \frac{2\pi\mathrm{i}}{L_{\rm eff}}x^{\text{new}}(j)
    \right],
    \quad
    z_j^q
    =
    \frac{\cosh\theta   e^{\mathrm{i}\frac{2\pi qj}{L}}-\sinh\theta}
    {\cosh\theta-\sinh\theta   e^{\mathrm{i}\frac{2\pi qj}{L}}},
    \quad
    j=1,\ldots,L.
\label{eq:app-HS-sites}
\end{equation}
The second equality specifies the map without the inverse trigonometric functions in Eq.~\eqref{eq:app-deformed-coordinate}. Its $q$th-root branch is fixed continuously from $z_j=e^{\mathrm{i}\frac{2\pi j}{L}}$ at $\theta=0$, equivalently by the continuous, increasing branch of $x^{\text{new}}(x)$. Here $x$ denotes the original continuous spatial coordinate and $f_{q,\theta}(x)$ is the deformation envelope defined earlier. Since $\mathrm{d}x^{\text{new}}/\mathrm{d}x=1/f_{q,\theta}(x)>0$ at finite $\theta$, the points in Eq.~\eqref{eq:app-HS-sites} remain distinct and preserve their cyclic ordering on the unit circle. The original conformal-block branch can therefore be continued throughout the deformation without braiding the insertions.

Direct differentiation of Eq.~\eqref{eq:app-HS-sites}, together with primary-field covariance under the chiral action of $U_q(\theta)$, determines the deformation of the full IDMPS. Denoting the transformed state by $\ket{\Psi_{q,\theta}}$, one obtains
\begin{equation}
\begin{split}
    \ket{\Psi_{q,\theta}}
    \propto&
    \sum_{\{s_j\}}
    \prod_{j=1}^{L}
    \left[
    \frac{
    e^{\mathrm{i}\frac{2\pi (q-1)j}{L}}
    }{
    z_j^{q-1}
    \left(
    \cosh\theta-\sinh\theta   e^{\mathrm{i}\frac{2\pi qj}{L}}
    \right)^2
    }
    \right]^{h_{s_j}}
    \left\langle
    \prod_{j=1}^{L}\varphi_{s_j}(z_j)
    \right\rangle
    \ket{s_1,\ldots,s_L}.
\end{split}
\label{eq:app-deformed-IDMPS}
\end{equation}
For each site $j$, the factor in square brackets is the derivative of the transformed plane coordinate $z_j$ with respect to the original plane coordinate $e^{\mathrm{i}\frac{2\pi j}{L}}$. Its fractional power is fixed by the same continuous continuation from $\theta=0$ used in Eq.~\eqref{eq:app-HS-sites}. At $\theta=0$, this factor equals one and Eq.~\eqref{eq:app-deformed-IDMPS} reduces to Eq.~\eqref{eq:app-uniform-IDMPS}. Any additional factor common to all local configurations is absorbed into the normalization.

If all local primary fields have the same chiral conformal weight, denoted by $h$, so that $h_{s_j}=h$ for every allowed local state $s_j$, the product of the Jacobian factors in Eq.~\eqref{eq:app-deformed-IDMPS} depends only on the insertion sites and not on the configuration $\{s_j\}$. It therefore cancels upon normalization, and the deformation reduces to the direct replacement of the uniform insertion points by the coordinates $z_j$:
\begin{equation}
    \ket{\Psi_{q,\theta}}
    \propto
    \sum_{\{s_j\}}
    \left\langle
    \prod_{j=1}^{L}\varphi_{s_j}(z_j)
    \right\rangle
    \ket{s_1,\ldots,s_L}.
\label{eq:app-equal-weight-IDMPS}
\end{equation}
Equations~\eqref{eq:app-deformed-IDMPS} and \eqref{eq:app-equal-weight-IDMPS} give the $q$-M\"obius deformation of a generic primary-field IDMPS. When the primary representations defining the conformal block contain null descendants, applying the corresponding null-field decoupling equations directly to the transformed correlator in Eq.~\eqref{eq:app-deformed-IDMPS} yields exact lattice annihilation conditions and hence positive-semidefinite parent Hamiltonians. Their interaction range and body order depend on the chosen CFT. We now make this construction explicit for the $\mathrm{SU}(2)_1$ Wess--Zumino--Witten theory, where the equal-weight simplification applies and the null-vector parent reduces to a long-range two-body Hamiltonian.

Let $L$ be even and specialize the local labels to $s_j=\pm1$, equal to twice the eigenvalue of $S_j^z$ at site $j$. The two values are the components of the spin-$1/2$ primary of the $\mathrm{SU}(2)_1$ theory and have the common chiral weight $h=1/4$. At the uniform roots of unity, the unique singlet chiral conformal block is the normalized Haldane--Shastry ground state $\ket{\mathrm{HS}_0}$, with normalization $\mathcal N_0$~\cite{CiracSierra2010,NielsenCiracSierra2011},
\begin{align}
    \ket{\mathrm{HS}_0}
    &= \frac{1}{\sqrt{\mathcal N_0}}
    \sum_{\{s_j=\pm1\}}
    \delta_{\sum_j s_j,0}
    \exp\left[
    \frac{\mathrm{i}\pi}{2}
    \sum_{j: \, \mathrm{odd}}(s_j-1)
    \right]\prod_{m<n}
    \left(
    e^{2\pi\mathrm{i}m/L}
    -
    e^{2\pi\mathrm{i}n/L}
    \right)^{s_ms_n/2}
    \ket{s_1,\ldots,s_L}, \nonumber \\
    \mathcal N_0
    &= \sum_{\{s_j=\pm1\}}
    \delta_{\sum_j s_j,0}
    \left|
    \prod_{m<n}
    \left(
    e^{2\pi\mathrm{i}m/L}
    -
    e^{2\pi\mathrm{i}n/L}
    \right)^{s_ms_n/2}
    \right|^2.
\label{eq:app-uniform-HS-state}
\end{align}
The Kronecker delta restricts the sum to the total-$S^z=0$ sector, and the phase factor is the Marshall sign. 

Because both local components have $h=1/4$, Eq.~\eqref{eq:app-equal-weight-IDMPS} applies. The normalized transformed conformal block, denoted by $\ket{\mathrm{HS}_{q,\theta}}$ with normalization $\mathcal N_{q,\theta}$, is therefore
\begin{equation}
\begin{split}
    \ket{\mathrm{HS}_{q,\theta}}
    =&
    \frac{1}{\sqrt{\mathcal N_{q,\theta}}}
    \sum_{\{s_j=\pm1\}}
    \delta_{\sum_j s_j,0}
    \exp\left[
    \frac{\mathrm{i}\pi}{2}
    \sum_{j: \, \mathrm{odd}}(s_j-1)
    \right]\prod_{i<j}(z_i-z_j)^{s_is_j/2}
    \ket{s_1,\ldots,s_L},\\
    \mathcal N_{q,\theta}
    =&
    \sum_{\{s_j=\pm1\}}
    \delta_{\sum_j s_j,0}
    \left|
    \prod_{i<j}(z_i-z_j)^{s_is_j/2}
    \right|^2.
\end{split}
\label{eq:app-deformed-HS-state}
\end{equation}
The fractional powers in Eq.~\eqref{eq:app-deformed-HS-state} use the same continuation from $\theta=0$ as Eq.~\eqref{eq:app-HS-sites}. Thus $z_j=e^{\mathrm{i}\frac{2\pi j}{L}}$, $\mathcal N_{q,0}=\mathcal N_0$, and $\ket{\mathrm{HS}_{q,0}}=\ket{\mathrm{HS}_0}$ exactly. At finite $L$, Eq.~\eqref{eq:app-deformed-HS-state} is the conformal-block lattice representative of the continuum state $U_q(\theta)\ket{0}$, where $\ket{0}$ denotes the CFT vacuum. The construction below gives an independent exact finite-size statement by proving that this normalized state is a ground state of an explicit microscopic long-range Hamiltonian.

For the $\mathrm{SU}(2)_1$ conformal block, let $z$ denote the complex coordinate of the holomorphic current $J^a(z)$, and let $S_j^a$ denote the spin-$1/2$ operator at site $j$. The indices $a,b,c$ take the values $x,y,z$, repeated $b,c$ indices are summed, and $s'_j$ labels the local spin state after the current acts. The singular part of the current-primary operator product expansion is
\begin{equation}
    J^a(z)\varphi_{s_j}(z_j)
    \sim
    \frac{1}{z-z_j}
    \sum_{s'_j}
    (S_j^a)_{s_js'_j}
    \varphi_{s'_j}(z_j).
\end{equation}
We write $\delta^{ab}$ for the Kronecker delta and $\varepsilon^{abc}$ for the Levi-Civita symbol. At level one, the descendant selected by $\delta^{ab}-\mathrm{i}\varepsilon^{abc}S_i^c$ is null~\cite{NielsenCiracSierra2011}. Inserting it into the conformal block and deforming the current contour from $z_i$ to all other insertions gives the Ward identity
\begin{equation}
    \sum_{j(\ne i)}
    \frac{
    S_j^a+\mathrm{i}\varepsilon^{abc}S_i^bS_j^c
    }{z_i-z_j}
    \ket{\mathrm{HS}_{q,\theta}}
    =0.
\label{eq:app-plane-null}
\end{equation}
This step uses only the residue in the displayed operator product expansion: the pole at $z_j$ supplies $S_j^b/(z_i-z_j)$, and contraction with the null-vector projector produces the two terms in Eq.~\eqref{eq:app-plane-null}.

Writing $\mathbf S_i=(S_i^x,S_i^y,S_i^z)$ for the spin vector at site $i$, the global Ward identity also gives
\begin{equation}
    \mathbf S_{\rm tot}\ket{\mathrm{HS}_{q,\theta}}=0,
    \quad
    \mathbf S_{\rm tot}=\sum_i\mathbf S_i,
\label{eq:app-HS-singlet}
\end{equation}
where $\mathbf S_{\rm tot}$ is the total spin. To put Eq.~\eqref{eq:app-plane-null} on the circle, multiply it by $2z_i$ and use
$2z_i/(z_i-z_j)=1+(z_i+z_j)/(z_i-z_j)$. The contribution from the first term vanishes because Eq.~\eqref{eq:app-HS-singlet} gives $\sum_{j(\ne i)}S_j^a\ket{\mathrm{HS}_{q,\theta}}=-S_i^a\ket{\mathrm{HS}_{q,\theta}}$ and, for spin $1/2$, $\mathrm{i}\varepsilon^{abc}S_i^bS_i^c=-S_i^a$. Define the real geometric coefficient $\kappa_{ij}$ by
\begin{equation}
    \frac{z_i+z_j}{z_i-z_j}
    =
    -\mathrm{i}\kappa_{ij},
    \quad
    \kappa_{ij}
    =
    \cot\left[
    \frac{
    \pi\bigl(
    x^{\text{new}}(i)-x^{\text{new}}(j)
    \bigr)
    }{L_{\rm eff}}
    \right].
\label{eq:app-HS-kappa}
\end{equation}
For each site $i$ and spin component $a$, define the operator $\mathcal A_i^a$ by
\begin{equation}
    \mathcal A_i^a
    =
    \sum_{j(\ne i)}
    \kappa_{ij}
    \left(
    S_j^a+\mathrm{i}\varepsilon^{abc}S_i^bS_j^c
    \right),
    \quad
    a=x,y,z.
\label{eq:app-HS-annihilator}
\end{equation}
Equation~\eqref{eq:app-plane-null} then implies
$\mathcal A_i^a\ket{\mathrm{HS}_{q,\theta}}=0$ for every site $i$ and every component $a$.

Define the null-vector parent Hamiltonian $H_{\rm null}$ by
\begin{equation}
    H_{\rm null}
    =
    \frac{1}{9}
    \sum_{i,a}
    \mathcal A_i^{a\dagger}\mathcal A_i^a,
\label{eq:app-HS-null-parent}
\end{equation}
where the sum runs over all sites $i$ and spin components $a$. This operator is positive semidefinite and annihilates $\ket{\mathrm{HS}_{q,\theta}}$. Writing $\mathbf 1$ for the identity operator and using $b'$ and $c'$ as additional Cartesian spin-component indices, the spin-$1/2$ identities
$S_i^aS_i^b=\delta^{ab}/4+\mathrm{i}\varepsilon^{abc}S_i^c/2$ and
$\sum_a\varepsilon^{abc}\varepsilon^{ab'c'}=\delta^{bb'}\delta^{cc'}-\delta^{bc'}\delta^{cb'}$
reduce its expansion to
\begin{equation}
    H_{\rm null}
    =
    \frac18
    \sum_{i\ne j}
    \kappa_{ij}^2 \mathbf 1
    +
    \frac16
    \sum_{i\ne j}
    \left(
    \kappa_{ij}^2
    +
    \sum_{k(\ne i,j)}
    \kappa_{ki}\kappa_{kj}
    \right)
    \mathbf S_i\cdot\mathbf S_j.
\label{eq:app-HS-null-expanded}
\end{equation}
For three distinct sites, the cotangent subtraction formula gives
\begin{equation*}
    \kappa_{ki}\kappa_{kj}
    =
    \kappa_{ij}(\kappa_{ki}-\kappa_{kj})-1.
\end{equation*}
Summing this identity over $k\ne i,j$ yields
\begin{equation}
    \sum_{k(\ne i,j)}
    \kappa_{ki}\kappa_{kj}
    =
    2\kappa_{ij}^2-L+2
    -
    \kappa_{ij}
    \left(
    \sum_{k(\ne i)}\kappa_{ik}
    -
    \sum_{k(\ne j)}\kappa_{jk}
    \right).
\label{eq:app-HS-cot-identity}
\end{equation}
Substituting Eq.~\eqref{eq:app-HS-cot-identity} into Eq.~\eqref{eq:app-HS-null-expanded}, using
\begin{equation*}
    1+\kappa_{ij}^2
    =
    \frac{1}{
    \sin^2\left[
    \pi\bigl(
    x^{\text{new}}(i)-x^{\text{new}}(j)
    \bigr)/L_{\rm eff}
    \right]
    },
\end{equation*}
and isolating the total-spin term through
$\sum_{i\ne j}\mathbf S_i\cdot\mathbf S_j
=\mathbf S_{\rm tot}^2-3L/4$,
one obtains the non-local two-body Hamiltonian
\begin{align}
    \widetilde H_{\mathrm{HS},q}(\theta)
    =&
    \sum_{i\ne j}
    \left\{
    \begin{aligned}
    &
    \frac{1}{
    4\sin^2\left[
    \frac{
    \pi\bigl(
    x^{\text{new}}(i)-x^{\text{new}}(j)
    \bigr)
    }{L_{\rm eff}}
    \right]
    }-
    \frac{\kappa_{ij}}{12}
    \left(
    \sum_{k(\ne i)}\kappa_{ik}
    -
    \sum_{k(\ne j)}\kappa_{jk}
    \right)
    \end{aligned}
    \right\}
    \left(
    2\mathbf S_i\cdot\mathbf S_j-\frac12
    \right).
\label{eq:app-deformed-HS-Hamiltonian}
\end{align}
Here $\widetilde H_{\mathrm{HS},q}(\theta)$ denotes the exact deformed Haldane--Shastry parent Hamiltonian. All sums over $i\ne j$ are ordered, so each physical pair is counted twice. Define the scalar $E_{\mathrm{HS},q}(\theta)$ by
\begin{align}
    E_{\mathrm{HS},q}(\theta)
    =&
    \sum_{i\ne j}
    \left\{
    \begin{aligned}
    &-
    \frac{1}{
    4\sin^2\left[
    \frac{
    \pi\bigl(
    x^{\text{new}}(i)-x^{\text{new}}(j)
    \bigr)
    }{L_{\rm eff}}
    \right]
    }+
    \frac{\kappa_{ij}}{24}
    \left(
    \sum_{k(\ne i)}\kappa_{ik}
    -
    \sum_{k(\ne j)}\kappa_{jk}
    \right)
    \end{aligned}
    \right\}
    -\frac{L}{4}.
\label{eq:app-deformed-HS-energy}
\end{align}
It completes the exact operator identity
\begin{equation}
    \widetilde H_{\mathrm{HS},q}(\theta)
    -
    E_{\mathrm{HS},q}(\theta)\mathbf 1
    =
    H_{\rm null}
    +
    \frac{L+1}{6}\mathbf S_{\rm tot}^2.
\label{eq:app-HS-positive-identity}
\end{equation}
Both terms on the right-hand side are positive semi-definite. Equations~\eqref{eq:app-HS-annihilator} and \eqref{eq:app-HS-singlet} show that both annihilate $\ket{\mathrm{HS}_{q,\theta}}$. Hence this normalized conformal-block state is an exact ground state of Eq.~\eqref{eq:app-deformed-HS-Hamiltonian}, with energy $E_{\mathrm{HS},q}(\theta)$ given by Eq.~\eqref{eq:app-deformed-HS-energy}. The argument establishes the ground-state property but does not require, or claim, uniqueness.

For $\theta=0$, one has
$\sum_{k(\ne i)}\cot[\pi(i-k)/L]=0$.
The second term in braces in Eq.~\eqref{eq:app-deformed-HS-Hamiltonian} therefore vanishes, and the first term reduces, up to the stated ordered-pair convention and an additive constant, to the uniform inverse-square Haldane--Shastry interaction.

No ground-state solver is needed to evaluate the overlap: the coefficients of $\ket{\mathrm{HS}_0}$ and $\ket{\mathrm{HS}_{q,\theta}}$ are given explicitly by Eqs.~\eqref{eq:app-uniform-HS-state} and \eqref{eq:app-deformed-HS-state}, and their scalar product can be summed directly over the neutral spin configurations. For $q=1$, the map is a global M\"obius transformation. Covariance of the primary-field correlator changes all spin configurations by the same factor, which cancels upon normalization. Consequently
$\lvert\langle\mathrm{HS}_0|\mathrm{HS}_{1,\theta}\rangle\rvert=1$
at every finite $L$. For $q \geq 2$ and $\theta \neq 0$, $|\mathrm{HS}_{q,\theta}\rangle$ differ from $|\mathrm{HS}_0\rangle$, and Eqs.~\eqref{eq:app-deformed-HS-state}--\eqref{eq:app-HS-positive-identity} establish an exact parent Hamiltonian for $|\mathrm{HS}_{q,\theta}\rangle$.

To test the CFT overlap prediction quantitatively, we evaluate the scalar product $\lvert\langle\mathrm{HS}_0|\mathrm{HS}_{q,\theta}\rangle\rvert$ directly from the explicit wave-function coefficients. 
For $q=2$, we extract the effective central charge through Eq.~\eqref{eq:app-CFT-overlap}. 
As shown in Fig.~\ref{fig:HS}, $c_{\rm eff}$ approaches the $\mathrm{SU}(2)_1$ value $c=1$ with increasing system size for all deformation strengths considered. 
Finite-size fits of the form $c_{\rm eff}(L)=c_\infty+aL^{-p}$ give extrapolated values close to unity. 
These results provide a direct long-range realization of the overlap construction using the analytically deformed conformal-block states, without numerically solving a deformed Hamiltonian.

\begin{figure}
    \centering
    \includegraphics[width=0.4\linewidth]{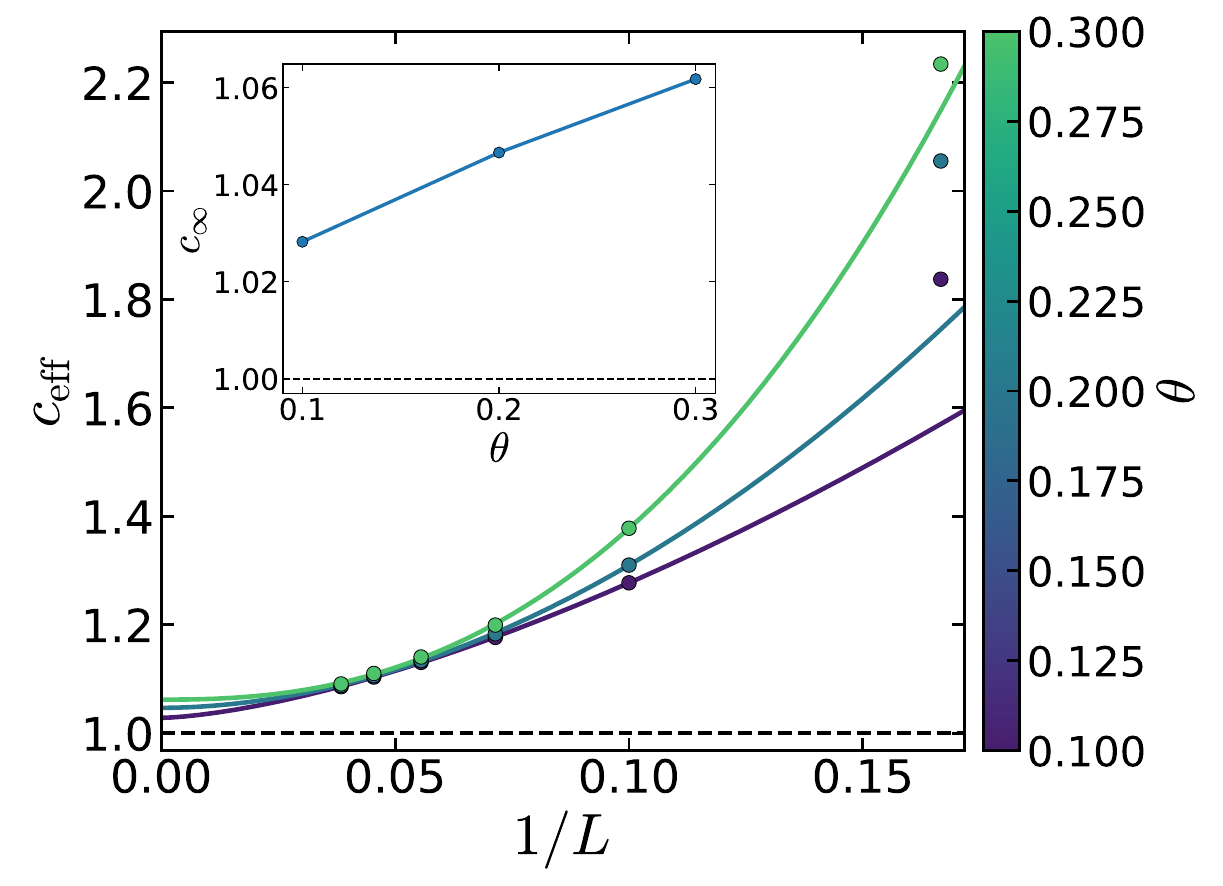}
    \caption{Central-charge extraction from the analytically deformed Haldane--Shastry state. 
    The effective central charge $c_{\rm eff}$, obtained from the overlap between the uniform and $q=2$ deformed conformal-block states, is plotted as a function of $1/L$ for different deformation strengths $\theta$. 
    Markers denote results for $L=6,10,14,18,22,26$, while solid curves show finite-size fits of the form $c_{\rm eff}(L)=c_\infty+aL^{-p}$ using the five largest system sizes. 
    The dashed horizontal line indicates the $\mathrm{SU}(2)_1$ CFT prediction $c=1$. 
    The inset shows the extrapolated values $c_\infty$ as a function of $\theta$.
    }
    \label{fig:HS}
\end{figure}
\end{widetext}

\end{document}